%% file: main.tex
\documentclass[a4paper,11pt]{article}
\usepackage{jheppub}

\usepackage{amsmath}
\usepackage{gensymb}
\usepackage{amsfonts}
\usepackage{comment}
\usepackage{amssymb}
\usepackage{mathrsfs}
\usepackage{graphicx}
\usepackage{color}
\usepackage{multirow}
\usepackage{bm}
\usepackage{tabularx}
\usepackage{subcaption}
\usepackage{blindtext}
\usepackage{wasysym}
\usepackage{makecell} 
\usepackage{hyperref}
\usepackage{float}
\hypersetup{colorlinks=true,allcolors=blue}
\usepackage{orcidlink}
\usepackage[normalem]{ulem}
\usepackage{lineno}
\usepackage[T1]{fontenc}
\usepackage{soul}
\usepackage[normalem]{ulem}
\usepackage{multirow}
\usepackage{array}
\usepackage{tikz-feynman}
\tikzfeynmanset{compat=1.1.0}
\newcolumntype{L}{>{\centering\arraybackslash}m{1.99cm}}

\newcommand{\vem}{$V_{e\mu}$~}
\newcommand{\vet}{$V_{e\tau}$~}
\newcommand{\vmt}{$V_{\mu\tau}$~}

\newcommand{\eq}{Eq.~}
\newcommand{\eqs}{Eqs.~}

\newcommand{\fig}{Fig.~}

\newcommand{\bi}{\begin{itemize}}
\newcommand{\ei}{\end{itemize}}

\newcommand{\lem}{$L_e - L_\mu$~}
\newcommand{\letau}{$L_e - L_\tau$~}
\newcommand{\lmt}{$L_\mu - L_\tau$~}
\newcommand{\ess}{ESSnuSB~}
\newcommand{\vab}{$V_{\alpha\beta}$~}
\newcommand{\dcp}{$\delta_{\rm CP}$~}
\begin{document}
 
\title{Probing Long-Range Forces in Neutrino Oscillations at the ESSnuSB Experiment
\\ \vspace{4mm}
{\small (ESSnuSB Collaboration)}}

\input{authors_copy.tex}

\authorlist

\abstract{Neutrino oscillations constitute an excellent tool to probe physics beyond the Standard Model. In this paper, we investigate the potential of the \ess experiment to constrain the effects of flavour-dependent long-range forces (LRFs) in neutrino oscillations, which may arise due to the extension of the Standard Model gauge group by introducing new $U(1)$ symmetries. 
Focusing on three specific $U(1)$ symmetries---$L_e - L_\mu$, $L_e - L_\tau$, and $L_\mu - L_\tau$, we demonstrate that \ess offers a favourable environment to search for LRF effects. Our analyses reveal that \ess can set 90\% confidence level bounds of $V_{e\mu} < 2.99 \times 10^{-14} \, \text{eV}$, $V_{e\tau} < 2.05 \times 10^{-14} \, \text{eV}$, and $V_{\mu\tau} < 1.81 \times 10^{-14} \, \text{eV}$, which are competitive to the upcoming Deep Underground Neutrino Experiment (DUNE). It is also observed that reducing the systematic uncertainties from $5\%$ to $2\%$ improves the \ess limits on $V_{\alpha\beta}$. Interestingly, we find limited correlations between LRF parameters and the less constrained lepton mixing parameters $\theta_{23}$ and $\delta_{\text{CP}}$, preserving the robustness of ESSnuSB's sensitivity to CP violation. Even under extreme LRF potentials ($V_{\alpha\beta} \gg 10^{-13} \, \text{eV}$), the CP-violation sensitivity and $\delta_{\text{CP}}$ precision remain largely unaffected. These results establish \ess as a competitive experimental setup for probing LRF effects, complementing constraints from other neutrino sources and offering critical insights into the physics of long-range forces.}


\maketitle

\section{Introduction}
The discovery of neutrino oscillations~\cite{Super-Kamiokande:1998kpq, SNO:2002tuh, Kajita:2016cak,  McDonald:2016ixn} has provided compelling evidence for physics beyond the Standard Model (SM), opening new avenues to explore new fundamental interactions and forces. The unique properties of neutrinos, including their elusive nature and tiny masses, make them an excellent probe to detect even the most subtle signatures of new physics. The neutrino experiments with their increasing precision are now sensitive to sub-leading effects due to potential non-standard interactions (NSIs), offering an indirect hint of new particles and forces not predicted by the SM.  

In the standard scenario, the interaction of neutrinos with matter is described by the so-called Mikheyev–Smirnov–Wolfenstein (MSW) mechanism, which results from coherent forward scattering of neutrinos with ambient matter~\cite{Wolfenstein:1977ue}. In this seminal paper, Wolfenstein also proposed the possibility of NSIs\footnote{In this manuscript, we will focus on new interactions mediated by vector bosons. There exists other forms of such interactions with different Lorentz structures \cite{Kopp:2007ne,Du:2020dwr,Gupta:2023wct,ESSnuSB:2023lbg,Denton:2024upc}, which physics signatures are however different from the ones discussed here.}, which have been extensively studied in the literature~\cite{Ohlsson:2012kf,Miranda:2015dra,Farzan:2017xzy,10.21468/SciPostPhysProc.2.001,Huitu:2016bmb,Chaves:2021kxe}.

In this work, we focus on another kind of such a new leptonic neutrino-matter interaction known as the long-range force (LRF), which may be flavour-dependent and mediated by light vector mediators~\cite{He:1990pn, Foot:1990mn,He:1991qd, Foot:1994vd,Dolgov:1999gk}. This is particularly intriguing since their effects can accumulate over astronomical distances, making them distinct from other NSIs. For instance, the matter content within astrophysical objects (Sun, Earth, Milky Way, etc.) can act as a source of LRF potential. These interactions significantly modify the probabilities of neutrino oscillations by introducing new potential terms in the Hamiltonian for neutrino propagation~\cite{Smirnov:2019cae}. Such interactions originate by extending the SM gauge group with additional anomaly-free $U(1)$ symmetries associated with lepton numbers $L_e, L_\mu, L_\tau$ and the baryon number $B$. We consider the three possible combinations of lepton flavours of symmetries~\cite{Pontecorvo:1967fh,Gribov:1968kq,Cirigliano:2005ck,Altarelli:2010gt}, for example, $L_e - L_\mu$, $L_e - L_\tau$, $L_\mu - L_\tau$. These symmetries are also important for generating neutrino masses~\cite{Asai:2017ryy,Asai:2018ocx,Lou:2024fvw}.  The constraints on the LRF parameters have already been obtained from solar~\cite{Grifols:2003gy,Bandyopadhyay:2006uh,Gonzalez-Garcia:2006vic}, atmospheric~\cite{Joshipura:2003jh} and astrophysical neutrinos~\cite{Bustamante:2018mzu,Agarwalla:2023sng} \footnote{Most works on LRFs in neutrino oscillations take into account specific models or mediator mass ranges. The only model-independent constraints on the LRF potentials from existing experiments come from high-energetic neutrinos observed at IceCube \cite{Agarwalla:2023sng} and of $\mathcal{O}(10^{-19} \, \, \mathrm{eV})$. These bounds are much tighter than the ones expected at terrestrial experiments due to energy-enhanced effects of LRFs. However, it is still worth exploring the bounds of accelerator experiments that employ a well-known and controlled neutrino beam.}.  In Ref.~\cite{Coloma:2020gfv}, a global analysis of three-flavour oscillation data has been performed in the presence of flavour-dependent long-range interactions. Furthermore, the effect of LRFs on long-baseline (LBL) neutrino experiments has been explored in Refs.~\cite{Chatterjee:2015gta,Khatun:2018lzs,Singh:2023nek,Mishra:2024riq}. 

 A key objective of present and future neutrino oscillation experiments \cite{ T2K:2017hed, T2K:2019bcf,NOvA:2019cyt,DUNE:2020jqi} is the precise determination of the leptonic CP-violating phase  $\delta_{\rm CP}$. The European Spallation Source (ESS) neutrino Super-Beam \ess \cite{ESSnuSB:2021azq} is a next-to-next-generation long-baseline neutrino oscillation experiment designed to achieve this goal. Located in Sweden, ESSnuSB will produce a high-intensity muon neutrino beam using a 5 MW proton beam from the upgraded ESS facility in Lund~\cite{Abele:2022iml, Alekou:2022emd}. The neutrinos will be detected by a water-Cherenkov detector situated 360 km away from Lund at the mine in Zinkgruvan. By focusing on the second oscillation maximum in the appearance probability $P_{\mu e}$, ESSnuSB is uniquely positioned to provide a precise measurement of $\delta_{\rm CP}$. Currently, ESSnuSB is at the stage of preparation of a second conceptual design report to be followed by the development of a technical design report. This will help to plan the construction and data collection at a later stage~\cite{ESSnuSB:2023ogw,ESSnuSB:2024tmn}. In the present work, we perform the first comprehensive study of the impact of long-range forces on the physics sensitivities of the ESSnuSB experiment. We derive bounds on the LRF potentials and the associated coupling parameters, comparing them with those achievable in the next-generation LBL experiments DUNE and T2HK. In addition, we investigate the effects of LRFs on the measurement of $\delta_{\rm CP}$ by ESSnuSB. Our analysis demonstrates that ESSnuSB's long baseline and high precision make it an ideal facility for probing the subtle effects of LRFs, offering sensitivity that surpasses those of some existing experiments.

 This paper is organized as follows. In Section \ref{sec:formalism}, we provide a brief overview of the theoretical framework of LRFs in neutrino oscillations, focusing on the three $U(1)$ symmetries under consideration. Then, in Section \ref{sec:simulation}, the description of the \ess experiment and other simulation details are provided. Next, in Section \ref{sec:prob_event_plots}, we compute the transition probabilities and generate the event plots in the presence of LRFs for ESSnuSB. In Section \ref{sec:bounds}, the sensitivity of the \ess experiment to constrain the LRF potentials and new coupling parameters are presented. Especially, Subsection \ref{sec:correlation} deals with some interesting correlations of LRF potentials with $\theta_{23}$ and $\delta_{\rm CP}$. Furthermore, in Section \ref{sec:CPV_sensitivity}, the impact of LRFs on the measurement of $\delta_{\rm CP}$ is discussed, which is followed by a precision study of CP violation (CPV) in Section \ref{sec:cpv_precision}. Finally, in Section \ref{sec:summaery}, we summarize our findings and conclusions.

\section{Theoretical formalism}
\label{sec:formalism}
Neutrino flavour transitions are significantly influenced by the interactions between neutrinos and the ambient matter as they propagate from the source to the detector. These interactions induce an effective potential in the Hamiltonian interaction~\cite{Wolfenstein:1977ue}. In standard scenario, neutrino-matter interactions occur through Charged Current (CC) and Neutral Current (NC) mechanisms.  While standard NC interactions are flavour-universal and do not impact neutrino oscillations, possible Beyond Standard Model (BSM) neutrino-matter interactions could introduce new potential terms that significantly alter neutrino propagation. Long-range forces are one such case, which may affect the measurements of neutrino oscillations in long-baseline experiments.

\subsection{Long-range forces from new $U(1)$ symmetries}
This can be envisaged by the extension of the SM gauge group $SU(3)_C\times SU(2)_L\times U(1)_Y$  with the minimal particle content by introducing the anomaly-free combination of the $U(1)$ symmetries $L_e, L_\mu, L_\tau$ and $B$ associated with the corresponding lepton numbers and baryon numbers. There are three possible lepton flavour combinations, for instance, $L_e - L_\mu$, $L_e - L_\tau$, \lmt, which can be gauged anomaly-free with the particles already present in the SM\footnote{In addition to these combinations, LRFs can also arise from other new U(1) symmetries, as discussed in Ref.~\cite{Agarwalla:2024ylc}. The corresponding \ess constraints on certain textures are provided in Appendix \ref{sec:appendix}.}. In principle, these extra symmetries cannot reproduce the neutrino observables \cite{Choubey:2024krp,Ibe:2025rwk}; however, with the addition of Higgs-like particles~\cite{Asai:2017ryy,Asai:2018ocx,Lou:2024fvw}, neutrino mixing and mass prediction can be reconciled. In addition, they induce a new flavour-dependent neutrino-matter interactions mediated by a new neutral gauge boson $Z^\prime$, and if the mediator is extremely light, the resulting forces might become significant over very large distances. The magnitude of the LRFs depends upon the matter contained within the radius $R_m\sim 1/m_{Z^\prime}$ which acts as a source of new potential.\\

The Lagrangian corresponding to the new interactions between a neutrino field $\nu_\alpha$ and a charged lepton field $l_\alpha$ mediated by a new gauge boson $Z^\prime$, for the combination of $U(1)$ symmetry $L_\alpha - L_\beta$,  is given by
\begin{align}
    \mathcal{L}_{Z^\prime}= g^\prime_{\alpha \beta}Z^\prime_{\rho}\left(\bar{l}_\alpha\gamma^\rho l_\alpha -\bar{l}_\beta\gamma^\rho l\beta + \bar{\nu}_\alpha \gamma^\rho \text{P}_L\nu_\alpha -\bar{\nu}_\beta\gamma^\rho \text{P}_L\nu_\beta\right)\, , 
    \label{eq:lagrangian}
\end{align}
where $g^\prime_{\alpha\beta}$ denote the dimensionless new gauge couplings and P$_L$ is the left-handed projection operator. It is to be noted that the (radiative) mixing~\cite{Babu:1997st,Joshipura:2019qxz} between $Z$ and $Z^\prime$ can also induce such long-range interactions, whose strengths are proportional to the couplings $g^\prime_{\alpha\beta}(\xi - \sin{\theta_w} \chi)$~\cite{Babu:1997st},  where the quantity $\chi$ is the kinetic mixing parameter between $Z$ and $Z'$~\cite{Holdom:1985ag,Babu:1997st}, the quantity $\xi$ is the rotation angle between mass and flavour bases of the gauge bosons and $\theta_w$ is the Weinberg angle. In Fig. \ref{fig:feynman}, we show the NC-like neutrino interactions mediated by $Z^\prime$ boson which can modify the oscillation probabilities; the left (right) diagram refers to the $L_e-L_\beta$ ($L_\mu-L_\tau$) case, see the following section for details.

\begin{figure}[h]\centering
\begin{tikzpicture}[scale=0.73, transform shape,every text node part/.style={align=center}]
 \begin{feynman}
    \vertex (a) {\Large\(\nu_\alpha\)};
    \node [below right=2.5cmof a] [dot](nodo1);
    \vertex [above =0.5cmof nodo1] (namenodo1){\Large\(g_{e\beta}^\prime\)};
    \vertex [above right=2.5cmof nodo1] (b) {\Large\(\nu_\alpha\)};
    \node [below=2.5cmof nodo1][dot] (nodo2);
    \vertex[below=0.5cmof nodo2] (namenodo2) {\Large\(g_{e\beta}^\prime\)};
    \vertex[below left=2.5cmof nodo2] (c) {\Large\(\ell_\alpha,u,d\)};
    \vertex[below right=2.5cmof nodo2] (d) {\Large\(\ell_\alpha,u,d\)};

    \diagram* {
      (a) -- [fermion,thick] (nodo1),
      (nodo1) -- [fermion,thick] (b),
      (nodo1) -- [boson,thick,edge label'=\Large{\(Z^\prime\)}] (nodo2),
      (c) -- [fermion,thick] (nodo2),
      (nodo2) -- [fermion,thick] (d),
    };
  \end{feynman}
\end{tikzpicture}
\hspace{2.5cm}
\begin{tikzpicture}[scale=0.73, transform shape,every text node part/.style={align=center}]
  \begin{feynman}
    \vertex (a) {\Large\(\nu_\alpha\)};
    \node [below right=2.5cmof a] [dot](nodo1);
    \vertex [above =0.6cmof nodo1] (namenodo1){\Large\(g_{\mu\tau}^\prime\)};
    \vertex [above right=2.5cmof nodo1] (b) {\Large\(\nu_\alpha\)};
    \node [below=2.4cmof nodo1][dot] (nodo2);
    \vertex[below=0.6cmof nodo2] (namenodo2) {\Large\(g_{Z}\)};
    \vertex[below left=2.5cmof nodo2] (c) {\Large\(\ell_\alpha,u,d\)};
    \vertex[below right=2.5cmof nodo2] (d) {\Large\(\ell_\alpha,u,d\)};
    \vertex [below=1.2cmof nodo1] (mid1);
    \vertex [right=0.3cmof mid1] (mid2) {{\Large{($\xi - \sin{\theta_w} \chi$)}}};

    \diagram*{
      (a) -- [fermion,thick] (nodo1),
      (nodo1) -- [fermion,thick] (b),
      (nodo1) -- [boson,thick,edge label'=\Large{$Z^\prime$},insertion=1] (mid1),
      (mid1) -- [boson,thick,edge  label'=\Large{$Z^{\phantom{\prime}}$}] (nodo2),
      (c) -- [fermion,thick] (nodo2),
      (nodo2) -- [fermion,thick] (d),
      };
  \end{feynman}
\end{tikzpicture}
    \caption{\footnotesize\textit{\textbf{Feynman diagrams of the new neutrino-matter interactions mediated by the new $Z^\prime$ neutral vector boson.}} The left diagram represents the interaction in the $L_e-L_\beta$ case, while the right diagram represents the interaction contributing to neutrino oscillations in the $L_\mu-L_\tau$ case, where the mixing between the $Z$ boson and the new $Z^\prime$ is needed. Here, $g_Z$ refers to the usual $Z$ boson coupling with leptons and quarks.}
    \label{fig:feynman}
\end{figure}

\subsection{Modified Hamiltonian due to long-range interaction potential}
\label{sec:osc}
Assuming the three new $U(1)$ gauge symmetries $L_e-L_\mu$, $L_e-L_\tau$ and $L_\mu-L_\tau$ that induce new neutrino-matter interactions, the effective Hamiltonian for neutrino propagation in the flavour basis is
\begin{align}
    H^{\rm eff} & = \frac{1}{2E}\left[U \begin{pmatrix} 
0 & 0 & 0 \\
0 & \Delta m^2_{21} & 0 \\
0 & 0 & \Delta m^2_{31}
\end{pmatrix} U^\dagger\right] \pm  V_{CC}
\begin{pmatrix} 
1 & 0& 0 \\
0 & 0& 0 \\
0 & 0 & 0
\end{pmatrix}  \pm V_{\alpha\beta}
\label{eq:Hamiltonian}.
\end{align}
In the above expression, $U$ is the standard Pontecorvo-Maki-Nakagawa-Sakata (PMNS) mixing matrix in vacuum, and $V_{CC} = \sqrt{2}G_F N_e$ is the usual matter potential term due to the CC interactions of neutrinos with matter\footnote{{NC interactions in neutrino propagation do not contribute to the oscillation probabilities in the usual three active neutrino framework. It is worth to mention that, in presence of sterile neutrinos or other forms of new physics, they could affect the matter-induced modifications of the vacuum neutrino oscillation probabilities~\cite{Ghoshal:2020hyo,Gehrlein:2024vwz,Denton:2022pxt,Agarwalla:2021owd,Coloma:2017ptb,Giarnetti:2021wur,Giarnetti:2024mdt,Berryman:2016szd}}.}. The signs of $V_{CC}$ and \vab are positive (negative) in the case of neutrino (antineutrino) oscillations. The neutrino energy is denoted by $E$, and $N_e$ is the electron number density. The contribution due to the long-range interaction is given by the new potential $V_{\alpha\beta}$ which, for the three different symmetries, can be written as 

\begin{equation}
 \label{equ:lrf_matrix}
 {V}_{\alpha\beta}
 =
 \left\{
  \begin{array}{ll}
   {\rm diag}(V_{e\mu}, -V_{e\mu}, 0), & {\rm for}~ L_e-L_\mu  \\
   {\rm diag}(V_{e\tau}, 0, -V_{e\tau}), & {\rm for}~ L_e-L_\tau \\
   {\rm diag}(0, V_{\mu\tau}, -V_{\mu\tau}), & {\rm for}~ L_\mu-L_\tau \\   
  \end{array}
 \right. \;.
\end{equation}
The specific form of the LRF Lagrangian generates a Yukawa-like potential with an interaction length inversely proportional to the mediator mass \cite{Bustamante:2018mzu,Wise:2018rnb}. Under transformations of the symmetry $L_e-L_\beta$, where $\beta = \mu, \tau$, this potential will be sourced by a population of electrons $N_e$ located at a distance $d$
from the neutrinos on Earth and is given as~\cite{Bustamante:2018mzu,Singh:2023nek}
\begin{equation}
\label{eq:pot_ve}
     V_{e\beta} = G_{e\beta}^{2} \frac{N_e}{4\pi d} e^{-m_{Z'} d}\,,
\end{equation}
where $G_{e\beta}$ is the effective coupling (which corresponds to $g'_{e\beta}$ in \eq (\ref{eq:lagrangian}), 
$m_{Z'}$ is the mass of new mediating gauge boson $Z'$. For $L_\mu - L_\tau$, the LRF is originated from the mixing between new gauge boson $Z'$ and the SM gauge boson $Z$~\cite{Heeck:2010pg,Joshipura:2019qxz}. In this case, assuming the Universe to be electrically neutral, the new potential experienced by neutrinos is only due to its interaction with $N_n$ number of neutrons which is given by~\cite{Heeck:2010pg}

\begin{equation}
 V_{\mu\tau}
 =
 G^2_{\mu\tau}\frac{e}{\sin{\theta_w}\cos{\theta_w}}\frac{N_n}{4\pi d} e^{-m_{Z'} d} \;,
 \label{equ:pot_vmt}
\end{equation}
where $e$ is the electric charge. For the \lmt symmetry, the effective coupling $G_{\mu\tau}$ is related to the coupling $g'_{\mu\tau}$ as  $G_{\mu\tau} = \sqrt{g'_{\mu\tau}(\xi - \sin{\theta_w} \chi)}$~\cite{Babu:1997st}.

It is worth mentioning that the structure of the new interaction potential is very similar to the standard matter potential except for the fact that in the former case, the mediator is extremely light ($Z'$), while in the latter case, the mediator is very heavy (the SM $Z$ boson). For this reason, LRFs can introduce new resonances in the transition probabilities at lower energies than the usual MSW resonance \cite{Wolfenstein:1977ue} such as \cite{Chatterjee:2015gta,Khatun:2018lzs} 
\begin{equation}
    E_{\rm{res}}=\frac{\Delta m_{31}^2 \cos2\theta_{13}}{2V_{CC}+3V_{e\beta}}\,,
\label{eq:Vebres}
\end{equation}
for the $L_e-L_\beta$ case. No resonances are expected in the $L_\mu-L_\tau$ scenario \cite{Agarwalla:2024ylc,Agarwalla:2021zfr}.
However, deriving expressions for neutrino oscillation probabilities in the presence of LRFs is cumbersome and not very enlightening. In some works, the ``effective'' mixing angles and mass-squared differences are computed using particular approximations and assumptions~\cite{Chatterjee:2015gta,Singh:2023nek,Mishra:2024riq}.
It should be noted that, at the Hamiltonian level, the effect of long-range forces is the same as the effect of flavour conserving vector NSIs \cite{Mishra:2024riq,Wise:2018rnb,Agarwalla:2021zfr,Giarnetti:2024mdt}. As it can be noted from analytical expansions presented in Refs.~\cite{Kopp:2007ne,Kikuchi:2008vq}, the flavour conserving vector NSI parameters appear in the oscillation probabilities as sub-leading effects in the $\nu_\mu\to\nu_e$ channel and at the first order in the $\nu_\mu\to\nu_\mu$ channel. However, given the presence of the new resonance in Eq.~(\ref{eq:Vebres}) in the $\nu_e$ appearance probability and since to be sensitive to LRF parameters we need $V_{\alpha\beta}$ of the order of the standard matter effect, the overall effects of LRFs cannot be fully understood from analytical expansions in small new physics parameters.
In this work, we discuss the LRF effects on the probabilities only numerically in Sec. \ref{sec:prob}.


\section{Simulation details of the ESSnuSB experiment}
\label{sec:simulation}
To generate the probability spectrum, analyze event rates, and perform sensitivity studies of \ess in the presence of LRFs, we employed the GLoBES software~\cite{Huber:2004ka, Huber:2007ji}. We introduced modifications to the probability engine to incorporate new potential terms due to the LRF as a new physics effect and then carried out numerical computations to obtain event rates and $\chi^2$ values. The experimental configuration and parameters for \ess used in our study are based on the \ess Conceptual Design Report \cite{Alekou:2022emd} and were therefore implemented in GLoBES. 

In particular, we considered a water Cherenkov far detector with a fiducial volume of 538~kt, positioned in the mine at Zinkgruvan, 360 km away from the neutrino source at ESS in Lund. A powerful linear accelerator (linac) will deliver $2.7\times 10^{23}$ protons on target per year, with a beam power of 5 MW and a proton kinetic energy of 2.5 GeV. We adopted updated neutrino fluxes, peaking at approximately 0.25 GeV, and applied updated migration matrices for event selection, as outlined in Refs.~\cite{Alekou:2022emd,ESSnuSB:2024yji}. The energy spectrum in the [0, 2.5] GeV range was divided into 50 bins for event calculations.

Our analyses included both the appearance ($\nu_\mu\to\nu_e$) and disappearance ($\nu_\mu\to\nu_\mu$) channels and their CP-conjugate transitions, and accounted for all the relevant backgrounds. We assumed systematic errors of $5\%$ for signals and $10\%$ for backgrounds unless otherwise stated. The total exposure time assumed for the far detector is 10 years, equally divided between 5 years of running the neutrino beam and 5 years for the antineutrino beam.
\section{Investigating LRFs at probability and event levels}\label{sec:prob_event_plots}
In this section, we first examine how the appearance and disappearance oscillations probabilities of muon neutrinos are influenced by the presence of a new interaction potential, $V_{\alpha\beta}$, sourcing the LRF at the \ess energies. Subsequently, we analyze the expected total event rates under the inclusion of LRFs in the theoretical framework. Unless stated otherwise, we adopt the best-fit values for the standard oscillation parameters from NuFIT 5.2~\cite{Esteban:2020cvm,NuFIT5.2}, which incorporate Super-Kamiokande atmospheric data and these parameters are summarized in  Table \ref{tab:t1}. For this analysis, we focus solely on the normal mass ordering (NO) for neutrinos, in line with the global fit preference for NO~\cite{Capozzi:2017ipn, Esteban:2018azc,deSalas:2020pgw,Capozzi:2021fjo,Gonzalez-Garcia:2021dve}, which might also be suggested by recent DESI-BAO cosmological results~\cite{Jiang:2024viw}.  
\begin{center}
\begin{table}
\centering
\begin{tabular}{|c |c| c|} 
 \hline
Oscillation parameters ($3\nu$) & Normal ordering (NO) \\ [0.5ex] 
 \hline\hline
$\theta_{12} (^{\circ})$ & $33.41^{+0.75}_{-0.72}$\\
 \hline
$\theta_{23} (^{\circ})$ & $42.2^{+1.1}_{-0.9}$\\
\hline
$\theta_{13} (^
{\circ})$ & $8.58^{+0.11}_{-0.11}$\\
\hline
$\delta_{\rm CP} (^{\circ})$ & $232^{+36}_{-26}$ \\
\hline
$\Delta m_{21}^2$ (eV$^2$) & $7.41^{+0.21}_{-0.20}\times 10^{-5}$ \\
\hline
$\Delta m_{31}^2$ (eV$^2$) & $+2.507^{+0.026}_{-0.027}\times 10^{-3}$ \\
\hline
\end{tabular}
\caption{ The best-fit value of the neutrino oscillation parameters in the standard three-flavour framework assuming normal mass ordering of neutrinos (NO). The values and their $1\sigma$ uncertainty intervals used in our calculations are taken from Ref.~\cite{Esteban:2020cvm}, which is the NuFit 5.2 data presented in 2022.}    \label{tab:t1}
\end{table}
\end{center}

\subsection{The $\nu_\mu\to\nu_e$ and $\nu_\mu\to\nu_\mu$ oscillation probabilities}
\label{sec:prob}
\begin{figure} 
\hspace*{-1cm}
     \centering
     \begin{subfigure}[b]{0.5\textwidth}
         \centering
         \includegraphics[width=\textwidth, height = 6cm]{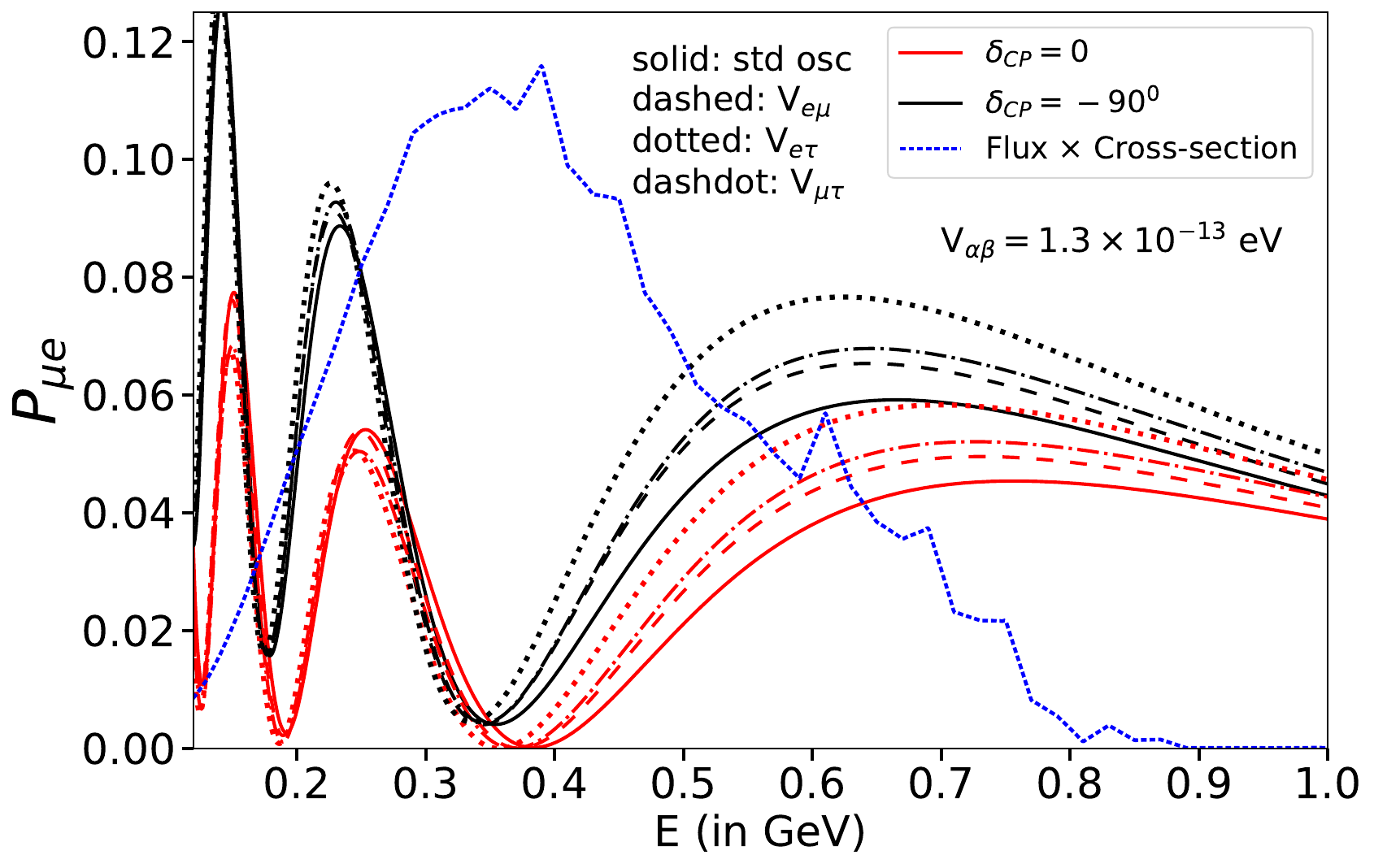}
         \caption{}
         
     \end{subfigure}
     \hfill
     \begin{subfigure}[b]{0.5\textwidth}
     \centering
     \includegraphics[width=\textwidth,height = 6cm]{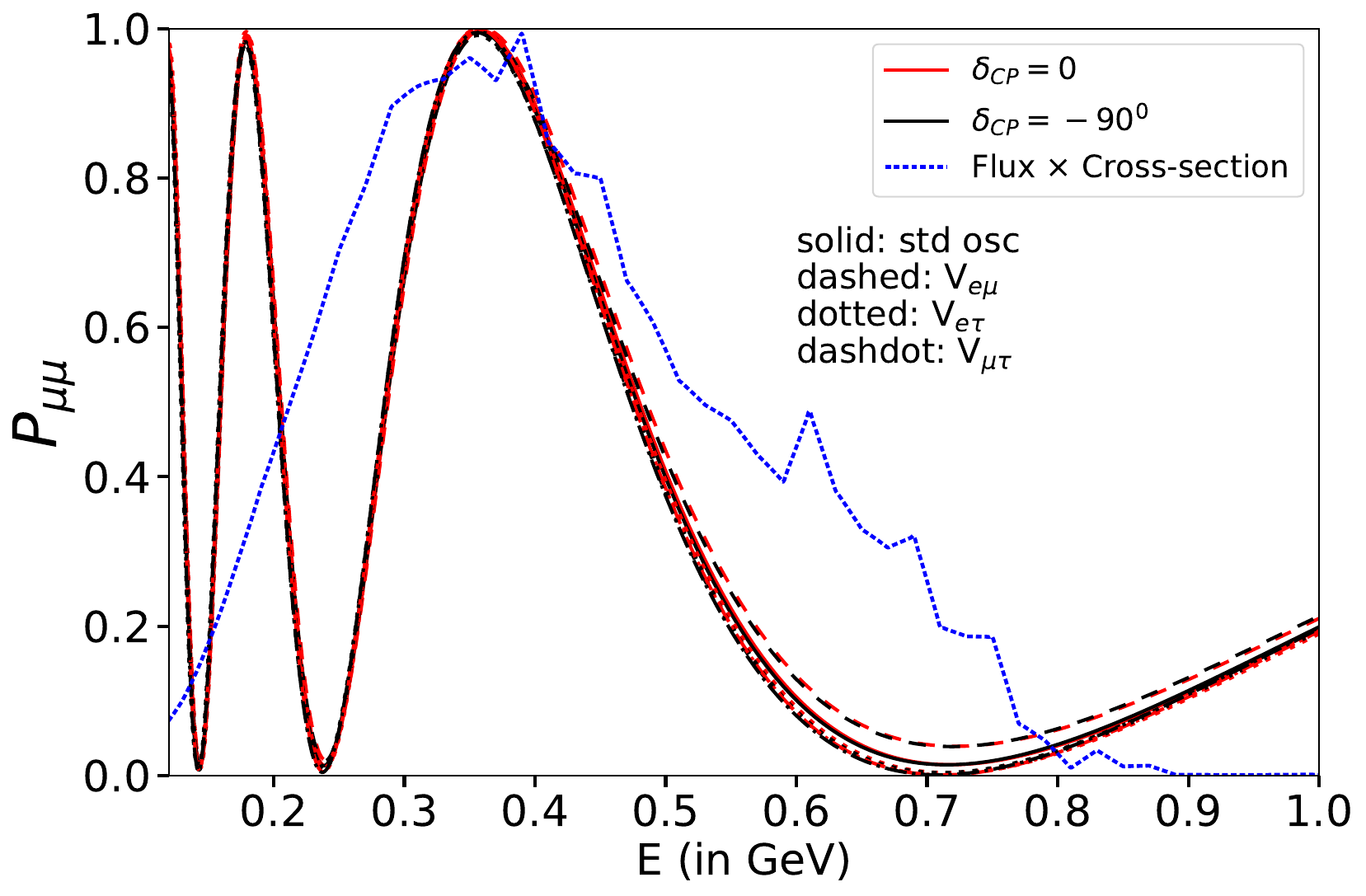}
     \caption{}
    
     \end{subfigure}
     \hfill
     \hspace*{-1cm}
    \begin{subfigure}[b]{0.5\textwidth}
         \centering
         \includegraphics[width=\textwidth,height = 6cm]{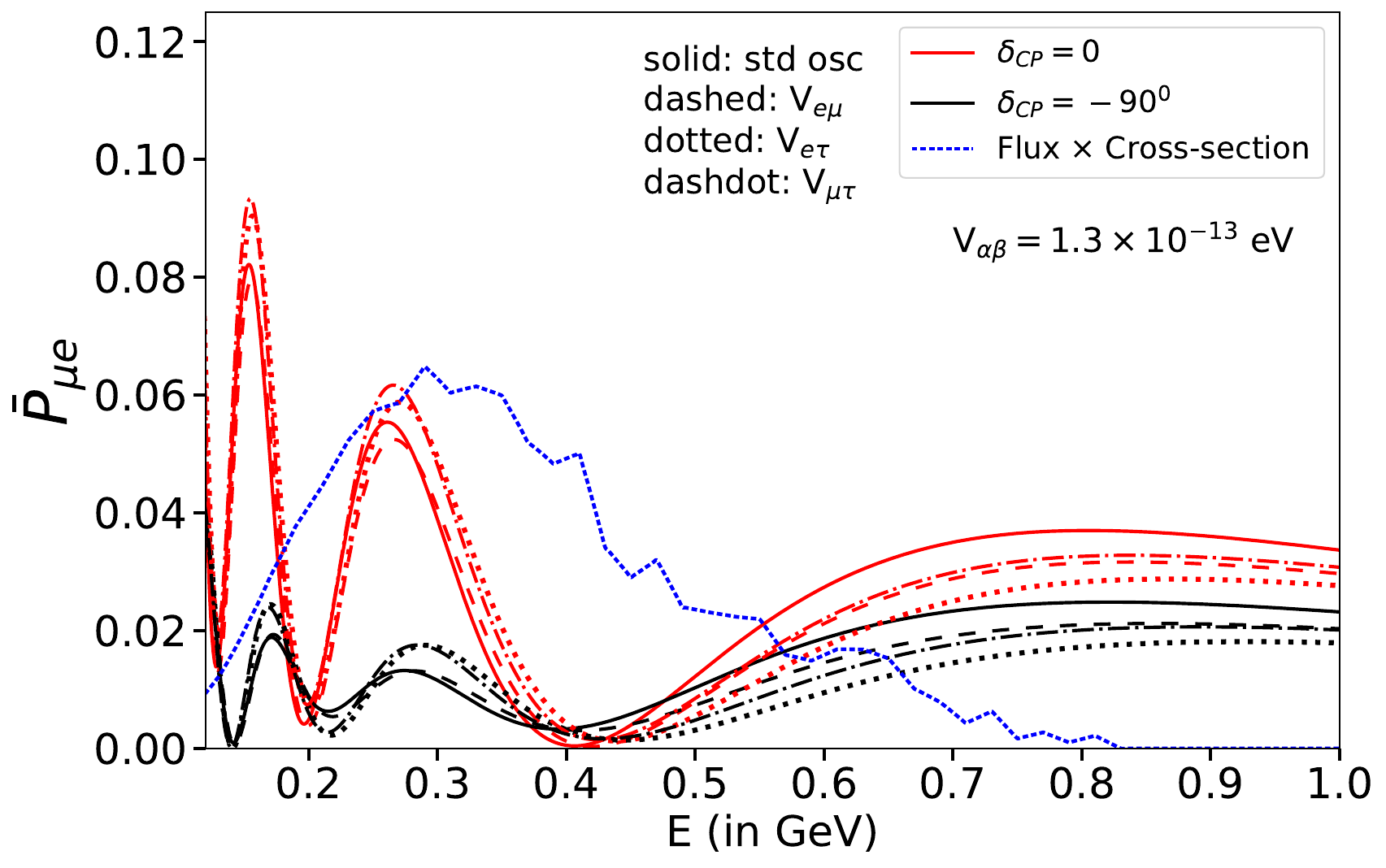}
         \caption{}
     \end{subfigure}
     \hfill
     \begin{subfigure}[b]{0.5\textwidth}
         \centering
         \includegraphics[width=\textwidth,height = 6cm]{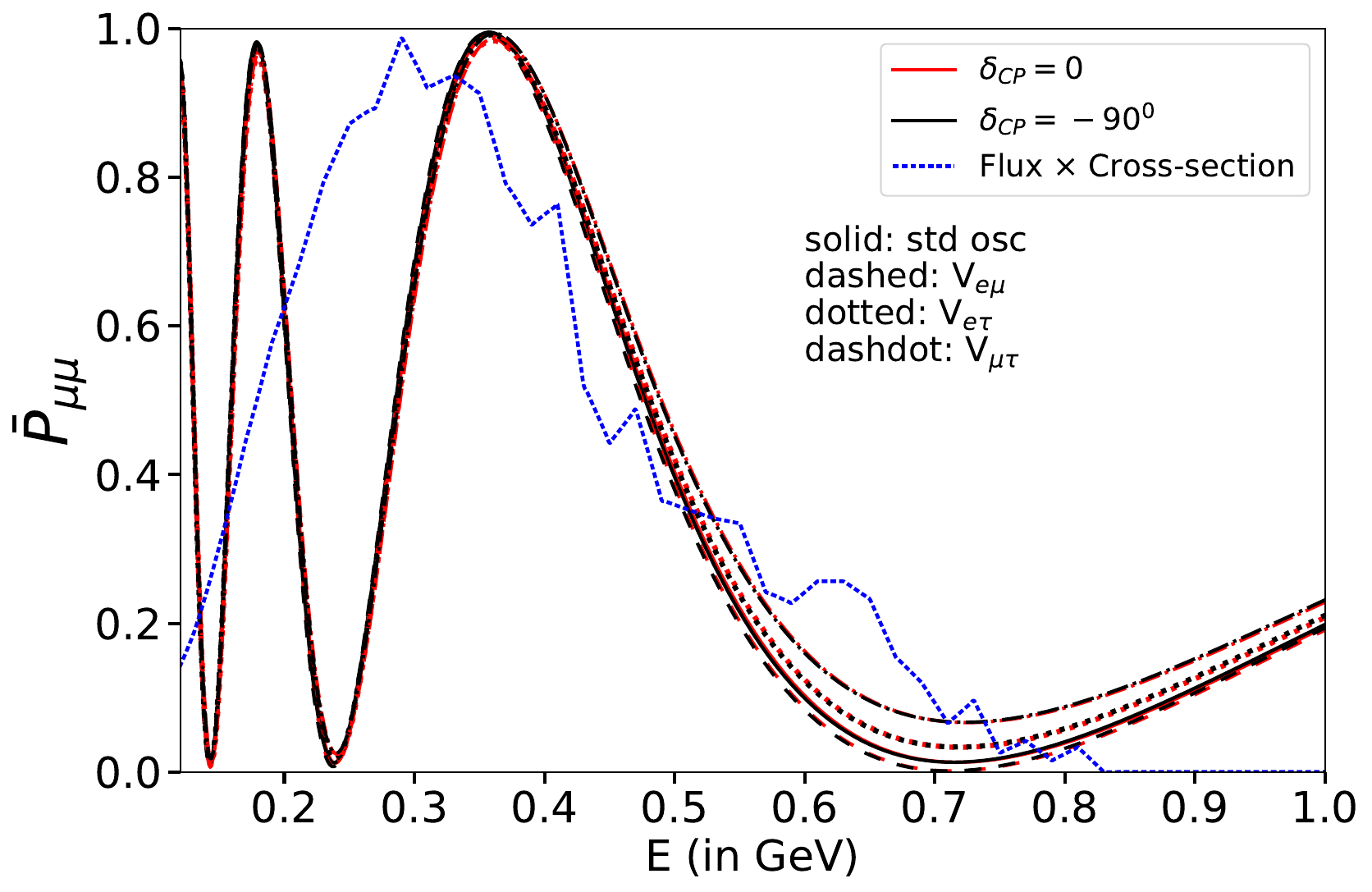}
         \caption{}
     \end{subfigure}
     \hfill
     \centering
     \caption{\footnotesize\textit{\textbf{Appearance (left panel) and disappearance (right panel) neutrino (top panel) and antineutrino (bottom panel) oscillation probabilities as functions of neutrino energy in the presence of LRF potentials, $V_{\alpha\beta} = 1.3\times 10^{-13}$ eV.}} The dashed, dotted and dashdot curves refer to the $L_e - L_\mu$, \letau and $L_\mu - L_\tau$ cases, respectively. The blue curve in each plot represents the $\text{flux}\times \text{cross-section}$ in the regions relevant for the \ess experiment.}
        \label{fig:prob}
\end{figure}

In \fig \ref{fig:prob}, we display the plots for neutrino oscillation probabilities, computed numerically in the presence of LRF potentials, $V_{\alpha\beta}$,  as a function of neutrino energy relevant for the \ess experiment. To show the impact of LRF potentials, we set their values, $V_{\alpha\beta} = 1.3\times10^{-13}$ eV, which is of the same order of magnitude as the standard matter potential \cite{Giarnetti:2024mdt,Singh:2023nek}.
The top (bottom) panel is presented for the neutrino (antineutrino) oscillation probability. The left (right) panel depicts the effect of $V_{\alpha \beta}$ on the appearance (disappearance) channel. In each panel, the solid curves denote the standard probabilities without $V_{\alpha\beta}$, while the dashed, dotted and dash-dotted curves refer to potentials, $V_{\alpha\beta}$,  corresponding to the three different symmetries, $L_e - L_\mu$, \letau and $L_\mu - L_\tau$, respectively.     
 Moreover, two extreme values for $\delta_{\rm CP}$ have been chosen, corresponding to the case of maximal CP violation ($\delta_{\rm CP}=-90^\circ$, black curve) and vanishing CP violation ($\delta_{\rm CP}=0^\circ$, red curve). To show the energy region relevant for the ESSnuSB experiment in each figure, we also superimpose the \ess flux multiplied by the charged current (CC) neutrino cross-section. 
 
 From \fig \ref{fig:prob} (top left), we observe that the neutrino appearance probability, $P_{\mu e}$, is enhanced around the first oscillation maximum in all three cases due to the presence of LRF potentials, whereas for antineutrino case, the appearance probability (bottom left), $\bar{P}_{\mu e}$, is suppressed. This is because the sign of LRF potential is flipped  ($V_{\alpha\beta}\to -V_{\alpha\beta}$) for the antineutrino case, similar to the standard matter potential. However, around the second oscillation maximum, the appearance probability increases for both the neutrino and antineutrino cases. For the neutrino appearance probability, $P_{\mu e}$, the first oscillation maximum also shifts towards lower energies for all three cases of LRF potentials, $V_{\alpha\beta}$. We also notice that the effects are more significant for the LRF potential $V_{e\tau}$ (dotted curve), whereas $V_{e\mu}$ (dashed curve) affects mildly. The disappearance channel, on the other hand, is less affected by $V_{\alpha\beta}$ compared to the appearance one for all three cases.  In particular, at the first oscillation minimum, the effect of $V_{e\mu}$ is larger for neutrinos (top right), while for antineutrinos (bottom right), the effect is more visible for $\mu\tau$ and $e\tau$ cases. However, given the much larger expected $\nu_\mu$ number of events at the far detector, the small disappearance probability modifications due to LRFs are crucial in constraining LRF potentials, $V_{\alpha\beta}$.

\subsection{Event rates in the presence of LRFs}
\begin{figure}[H]
     \centering
         \includegraphics[width=7.5 cm, height=7.3cm]{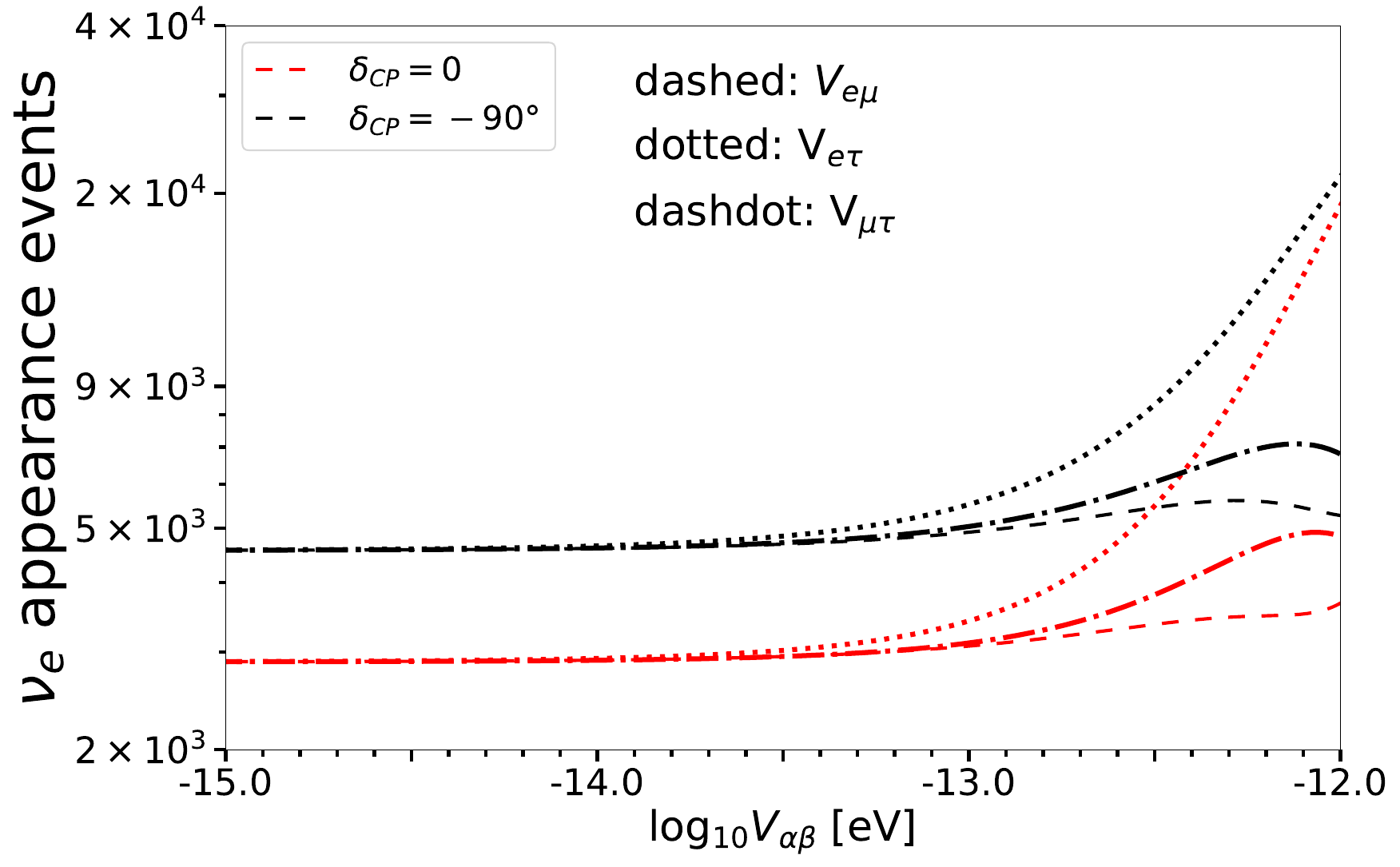}
         \includegraphics[width=7.5 cm, height=7.3cm]{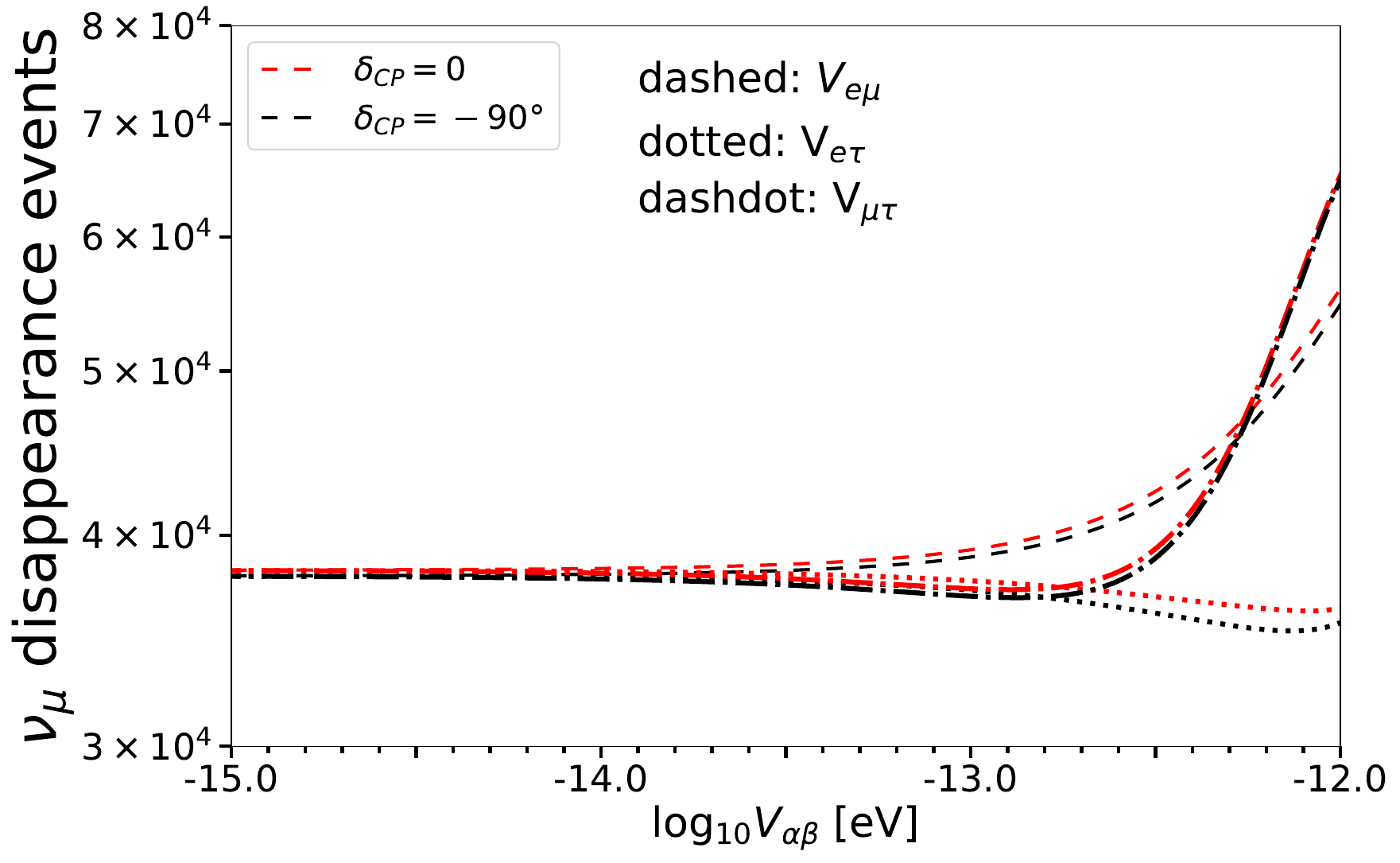}
\caption{\footnotesize\textit{\textbf{Total expected number of events is plotted as a function of LRF potential for the \ess experiment for two choices of $\delta_{\rm CP} = 0$ and $-90\degree$.}} Appearance events are shown in the left panel and disappearance events in the right panel.}
        \label{fig:event360}
\end{figure}

In order to make an initial guess about the limits that \ess would set on LRF parameters, $V_{\alpha\beta}$, we plot the total number of neutrino appearance (and disappearance) events as a function of the LRF parameter for 10 years of running, 5 in neutrino and 5 in antineutrino mode. The potential, $V_{\alpha\beta}$, is varied from $10^{-15}$ eV to $10^{-12}$ eV. The results are presented in Fig. \ref{fig:event360}, where the left panel is for the appearance of electron neutrino events and the right one refers to events corresponding to the disappearance of muon neutrinos. The black curves in each plot depict the case of maximal CP violation ($\delta_{\rm CP}=-90^\circ$), while red curves refer to the case of CP conservation ($\delta_{\rm CP}=0^\circ$). The features observed in the discussion of the probabilities can be directly translated into these plots. Indeed, in each case, we can observe a transition from the standard case (without LRFs) to the LRF-dominated case. The transition begins for values of LRF potentials, $V_{\alpha\beta}$, for which the correction in the standard probability due to the presence of LRFs overcomes the standard matter probability.
 A rough estimate of the constraint on \vab that is obtainable from the \ess experiment can be derived directly from \fig \ref{fig:event360}, \vab $\ll 10^{-13}$ eV; indeed, for larger potentials the expected number of events is much larger than the one expected in the case $V_{\alpha\beta}\to0$. However, the detailed $\chi^2$ analysis illustrated in the following section reveals stronger and more precise bounds on the LRF potentials. It can be observed from \fig \ref{fig:event360} that for $V_{e\tau}$ in the appearance case and $V_{e\mu}$ in the disappearance one, the number of events increases for both values of $\delta_{\rm CP}$. The $\nu_\mu$ disappearance events, however, decrease in the range $10^{-14}-10^{-13}$ eV, with \vab under transformations of the \letau and \lmt symmetries. 
         
    


\section{Constraint plots for ESSnuSB}
\label{sec:bounds}
In this section, we explore the capability of the \ess experiment to constrain the parameters of LRFs. The statistical analysis has been performed using a Poissonian $\chi^2$ function, defined as
\begin{equation}
    \chi^2(\Vec{\Lambda},b)=2\sum_{i=1}^n \left[(1+b)E_i-O_i+O_i\log\frac{O_i}{(1+b)E_i}\right]+\frac{b^2}{\sigma_b^2}\,,
\end{equation}
where $\Vec{\Lambda}$ represents the set of oscillation parameters needed to compute the rates, $\sigma_b$ is the normalization error, $n$ is the number of energy bins, $O_i$ are the observed rates and $E_i$ are the expected rates used for the fit. Systematic uncertainties are incorporated using the \textit{pull method} \cite{Huber:2002mx,Fogli:2002pt}, implemented in GLoBES with the nuisance parameter $b$. The significance of our results in terms of standard deviations ($\sigma$) has been obtained assuming the Wilk's theorem \cite{Wilks:1938dza}; for instance, for 1 d.o.f $\#\sigma = \sqrt{\Delta \chi^2}$.

\begin{figure}
\hspace*{-1.5cm}
\includegraphics[width=7.5cm,height=6.0cm]{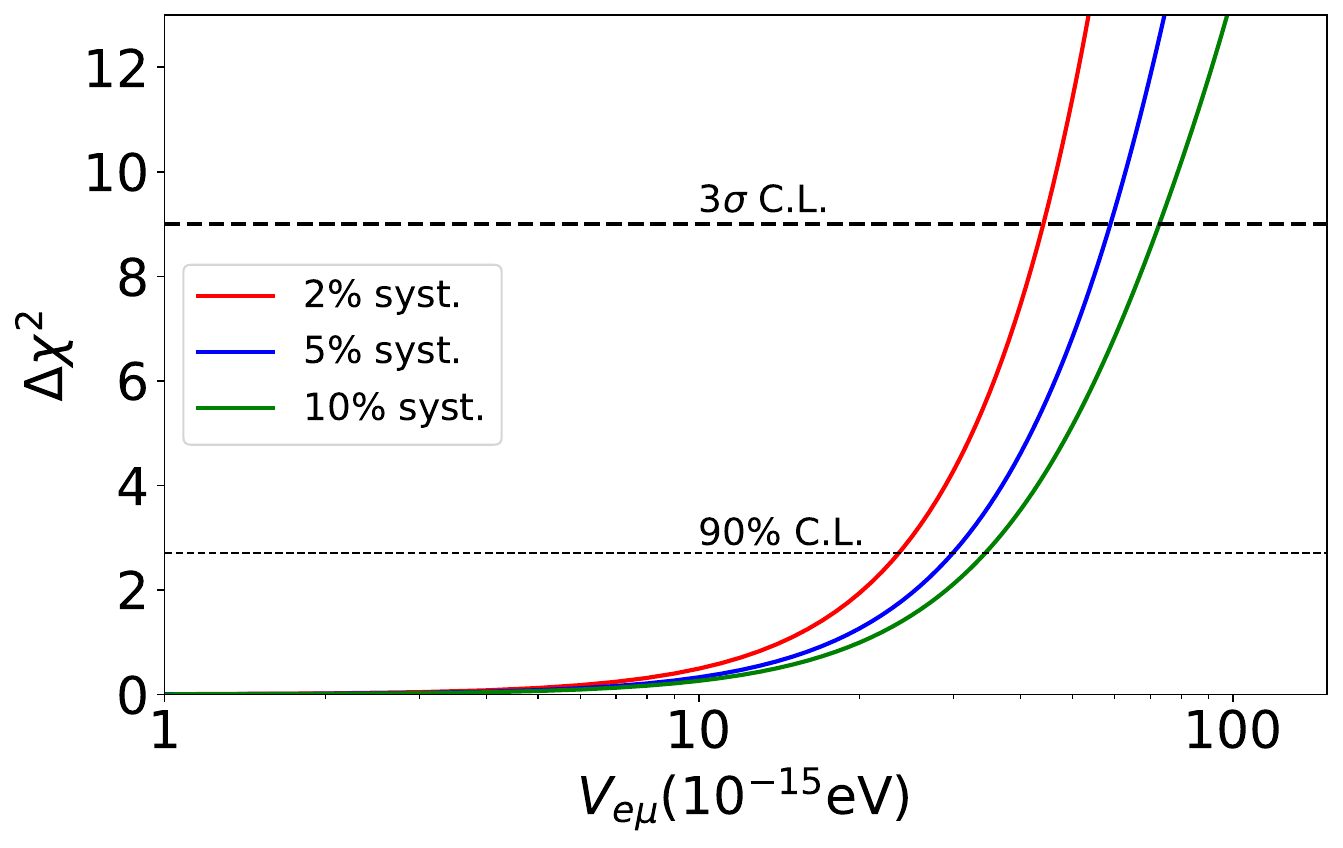}~~
\hspace*{0.5cm}
\includegraphics[width=7.5cm,height=6.0cm]{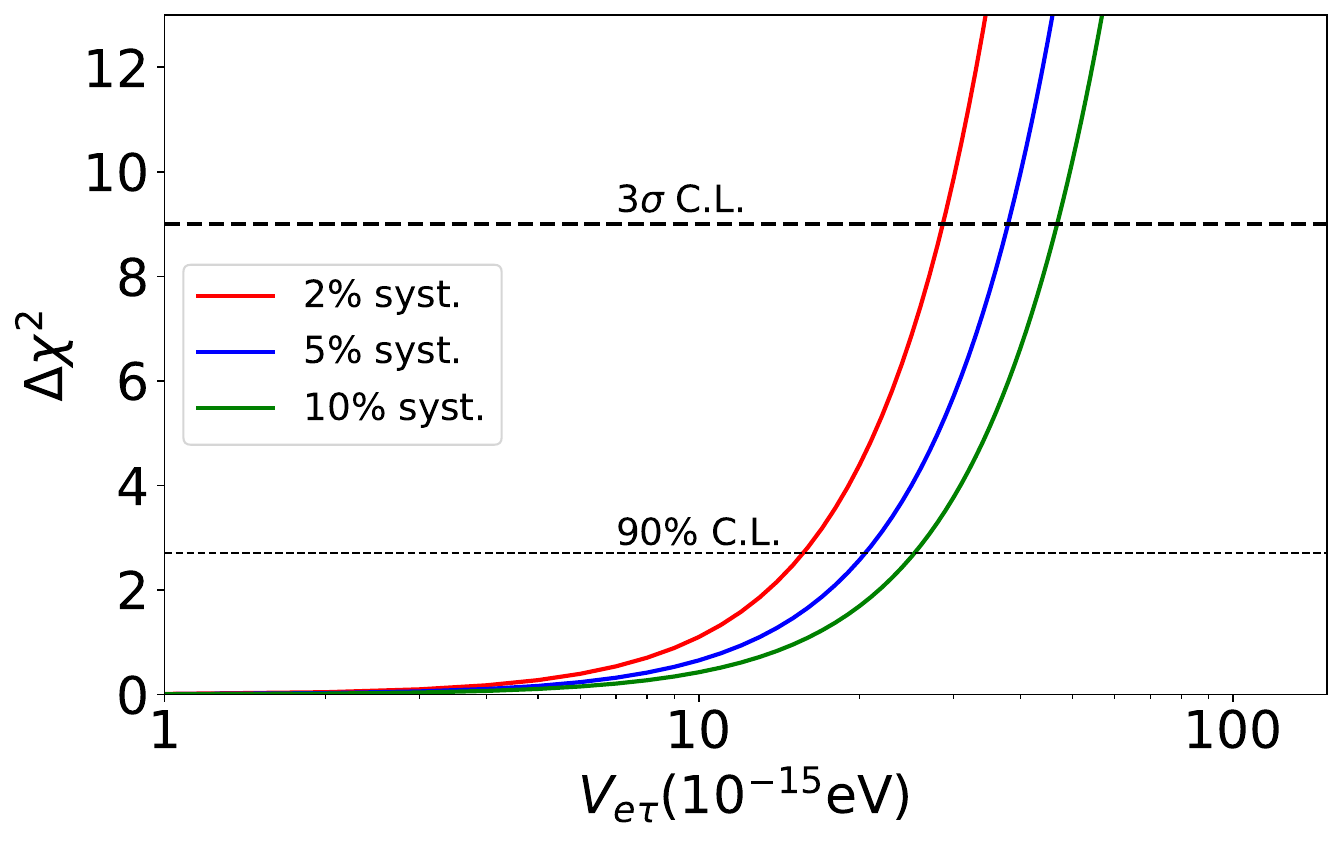}
\vskip 0.8cm
\centering \includegraphics[width=8cm,height=6.0cm]{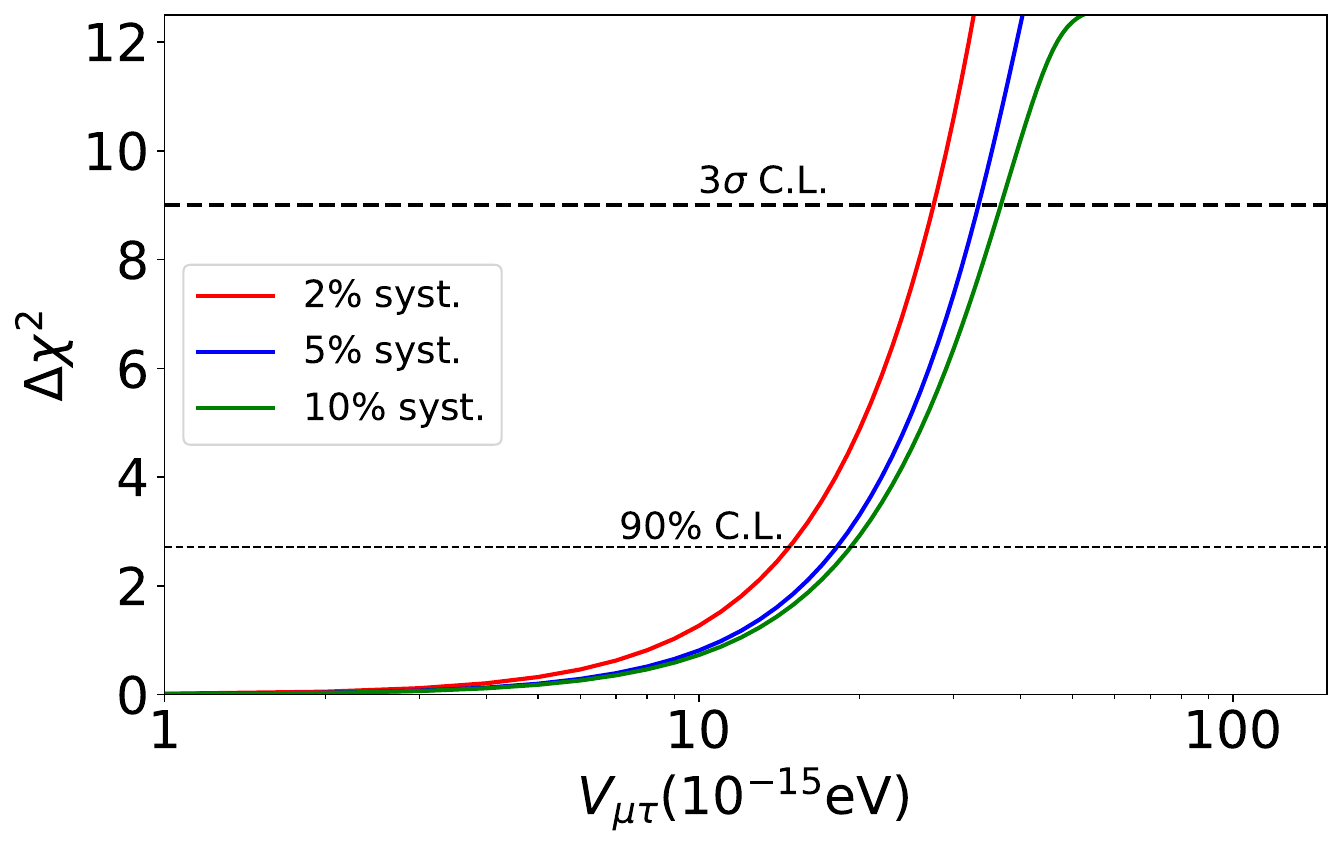}
\caption{\footnotesize \textit{\textbf{Sensitivity of \ess in constraining the LRF potential $V_{\alpha\beta}$.}} We consider normal mass ordering for neutrinos i.e., $\Delta m^2_{31} >0$. The red, blue and green colours correspond to the cases for $2\%, 5\%$ and $10\%$ systematic uncertainties, respectively.}
        \label{fig:chisq}
\end{figure}

In order to compute the bounds on $V_{\alpha\beta}$, we generate the true event spectrum using the hypothesis of no LRFs (i.e.~\vab(true) = 0) corresponding to standard three-flavour neutrino oscillations and fit the true data using the probabilities in the presence of LRFs. It should be noted that, while fitting the true data, only one LRF parameter is considered at a time in the test. This approach is justified as different potentials stem from distinct symmetries, independently affecting the oscillations. In all three cases of symmetries, we vary the potentials, \vab from $10^{-15}$ eV to $10^{-13}$ eV in the test. The marginalization has been performed over $\theta_{13}, \theta_{23}$ and $|\Delta m^2_{31}|$ by varying them within the uncertainty ranges reported in Table \ref{tab:t1}, while $\delta_{\rm CP}$ is scanned over its full $[-180\degree, 180\degree]$ range. We keep the two oscillation parameters, $\theta_{12}$ and $\Delta m^2_{21}$, fixed at their best-fit values~\cite{Esteban:2020cvm}. The results are displayed in \fig \ref{fig:chisq} where the one-dimensional $\Delta \chi^2$ is plotted as a function of LRF potentials  $V_{\alpha\beta}$.  The upper left (right) plot of \fig \ref{fig:chisq} gives the bound on the LRF parameter $V_{e\mu}~(V_{e\tau})$ while the lower plot displays the constraint on $V_{\mu\tau}$. We also show the results for different values of the normalization systematic uncertainty, namely 2\% (red curves), 5\% (blue curves) and 10\% (green curves). The 3$\sigma$ and 90\% C.L. bounds are summarized in Table \ref{tab:bound_table} for the standard 5\% systematics case along with the 2\% and 10\% systematics cases. The main results are that \ess in the nominal conditions (i.e. 5\% systematics) may be able to set the 90\% limits on $V_{e\mu}<2.99\times10^{-14}$ eV, $V_{e\tau}<2.05\times10^{-14}$ eV and $V_{\mu\tau}<1.81 \times 10^{-14}$ eV. Notably, a change between $10\%$ and $20\%$ in the bounds can also be observed by variations in systematic uncertainties, particularly when $V_{e\mu}$ and $V_{e\tau}$ are considered. The effect of systematics on \vmt is, on the other hand, less prominent.

\begin{figure}
\hspace*{-1.5cm}
\includegraphics[width=7.5cm,height=6.0cm]{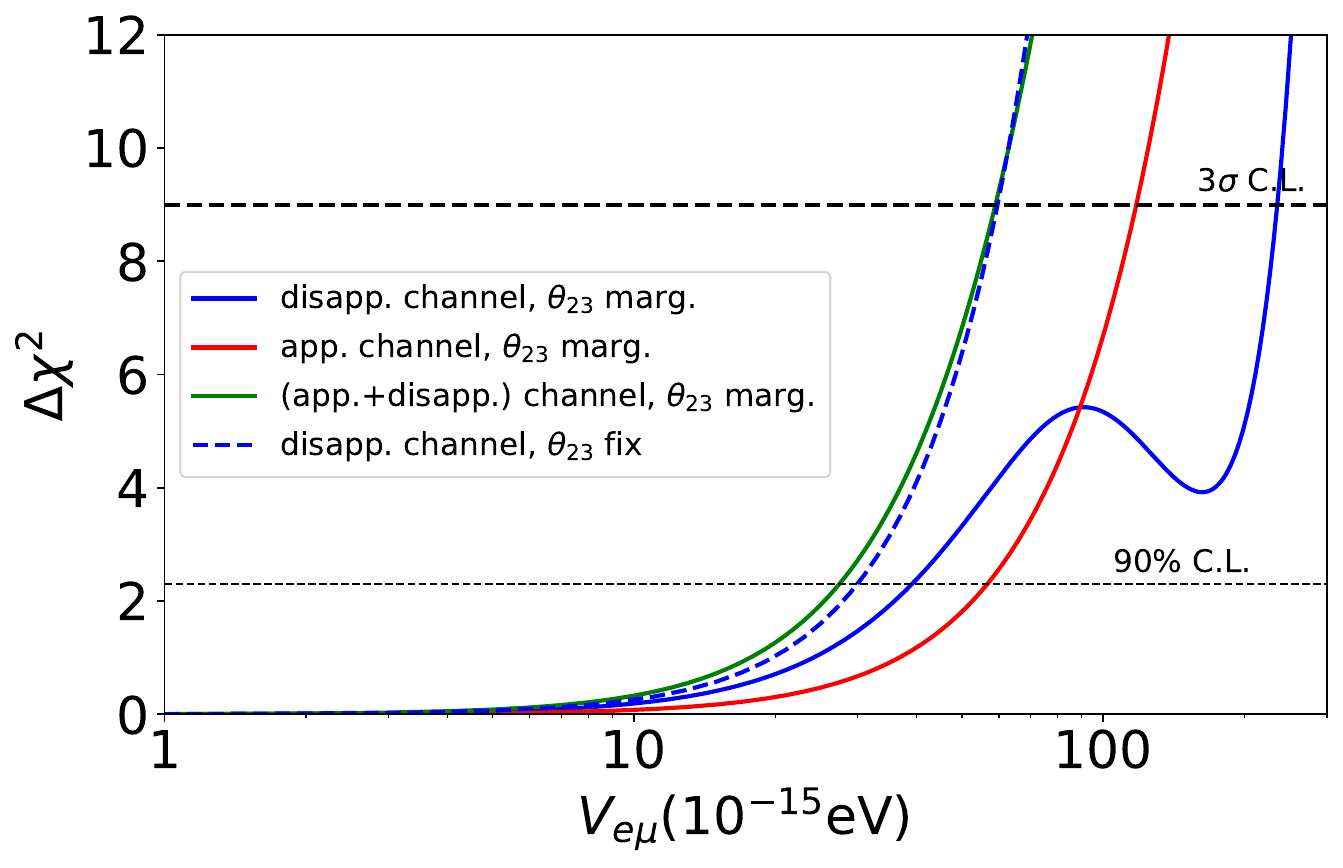}~~
\hspace*{0.5cm}
\includegraphics[width=7.5cm,height=6.0cm]{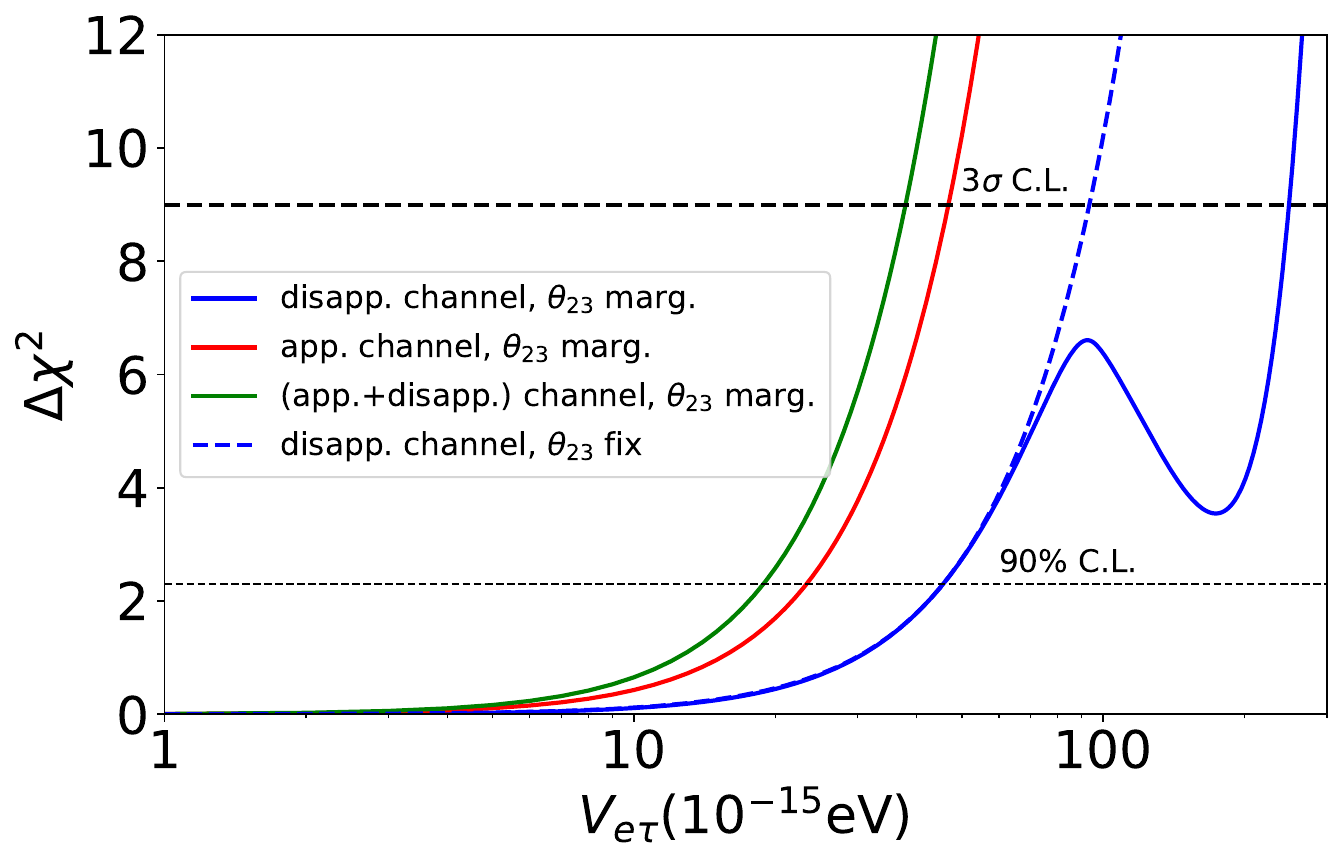}
\vskip 0.8cm
\centering \includegraphics[width=8cm,height=6.0cm]{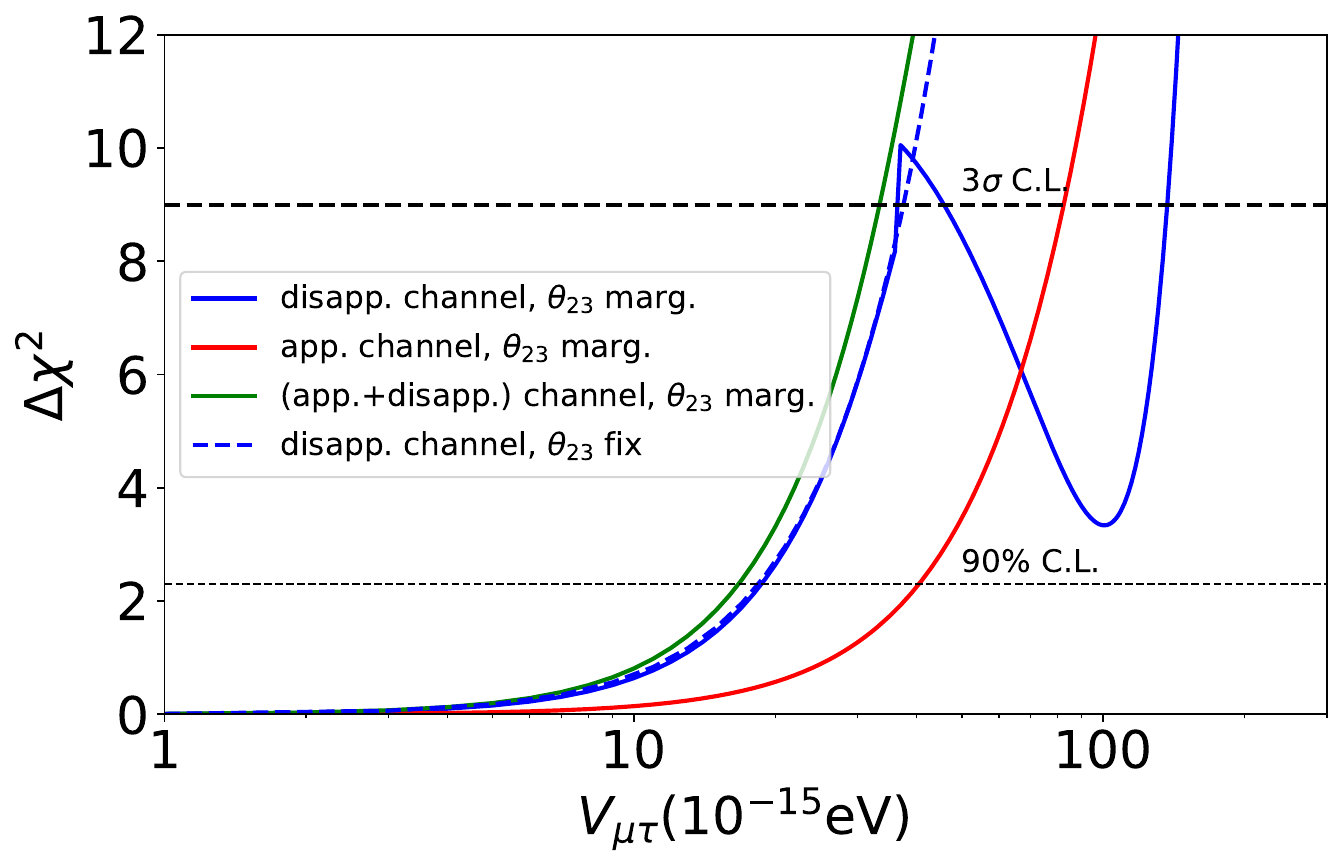}
\caption{\footnotesize \textit{\textbf{Significance of appearance and disappearance channels in computing the sensitivity of \ess to constrain the LRF potentials.}} Dashed curves show the effect of fixing $\theta_{23}$ to its best-value in the disappearance channel.} 
        \label{fig:chisq_channel}
\end{figure}

Before comparing the \ess limits on LRF potentials with other experimental bounds, let us try to understand the role of appearance and disappearance channels in constraining $V_{\alpha\beta}$. In \fig \ref{fig:chisq_channel}, we demonstrate how individual probability channels contribute for the \ess sensitivity towards LRF potentials, $V_{\alpha\beta}$. It is evident that for the \lem and \lmt symmetries the major sensitivity comes from the disappearance ($P_{\mu\mu}$) probability, whereas the appearance probability ($P_{\mu e}$) plays a major role to place a bound on the $V_{e\tau}$ potential corresponding to the \letau symmetry. This is also clear from the probability plots presented in \fig \ref{fig:prob}, where the effect of the $e \tau$ sector is more visible in the appearance probability ($P_{\mu e}$). Although from \fig \ref{fig:prob}, it seems that the $e\mu$ and $\mu\tau$ sectors affect both $P_{\mu e}$ and $P_{\mu\mu}$, however, due to the high statistics of $\nu_\mu$ the disappearance event numbers at the far detector, disappearance channel plays an important role in constraining \vem and $V_{\mu\tau}$. This explains why different oscillation channels are sensitive to different LRF potentials. In \fig \ref{fig:chisq_channel}, we notice a dip corresponding to the disappearance-only sensitivity curves (blue solid curves) for all three cases of the LRF potentials when a marginalization over $\theta_{23}$ is performed. Similar features are also observed in other works~\cite{Singh:2023nek,Mishra:2024riq}. This is because, in the disappearance probability, the octant of $\theta_{23}$ develops a degeneracy with the potential \vab picking up the wrong solution in the minimum $\chi^2$ calculation when marginalization is performed over $\theta_{23}$. The dip vanishes for the disappearance only case if we fix $\theta_{23}$ to its best-fit value while computing the $\chi^2$. Also, the dip disappears when we combine both the appearance and disappearance channels (green solid curves), while marginalizing over $\theta_{23}$, highlighting the importance of the appearance channel, which is less affected by the $\theta_{23}$ octant degeneracy.

In the context of LBL experiments, the most stringent foreseen $90\%$ C.L. limits on LRF potentials have been derived by simulating the future experiment P2SO~\cite{Mishra:2024riq} due to its longer baseline, whereas the bounds (at $90\%$ C.L.) from the simulations of  ``upcoming'' DUNE and T2HK experiments with the standard neutrino flux are given by~\cite{Singh:2023nek}:
\begin{align}\label{eq:dune_bounds}
    V_{e\mu}& < 1.46~ (3.45) \times 10^{-14} ~\text{eV}~~ [\text{DUNE~(T2HK)}]\,,\nonumber\\
    V_{e\tau} & < 1.03~ (3.43)\times 10^{-14} ~\text{eV}~~ [\text{DUNE~(T2HK)}]\,,\nonumber\\
    V_{\mu\tau}& < 0.67~ (1.84)\times 10^{-14} ~\text{eV}~~ [\text{DUNE~(T2HK)}].
\end{align}
It is worth mentioning that with a high-energy neutrino flux, the DUNE bounds on \vab might become weaker as shown in Ref.~\cite{Giarnetti:2024mdt}. 
\begin{table}[H]
\centering
\begin{tabular}{|L|c|c|c|c|c|c|}
\hline
\multirow{3}{*}{}{LRF Potential} (in eV) & \multicolumn{3}{c|}{$3\sigma$ C.L. } & %
    \multicolumn{3}{c|}{$90\%$ C.L.}\\
\cline{2-7}

&$2\%$ syst.  & $5\%$ syst. & $10\%$ syst. &$2\%$ syst.  & $5\%$ syst. & $10\%$ syst.\\
\hline
$V_{e\mu} (\times 10^{-14})$ & $4.41$  & $5.89$  & $7.28$  &$2.37$ &$2.99$ &$3.44$ \\
\hline
$V_{e\tau} (\times 10^{-14})$ &$2.86$ &$3.79$ &$4.68$ &$1.57$ &$2.05$ &$2.54$ \\
\hline
$V_{\mu\tau} (\times 10^{-14})$  &$2.75$ & $3.34$&$3.67$ &$1.48$ &$1.81$ & $1.92$\\
\hline
\end{tabular}
\caption{Constraints on the LRF potential $V_{\alpha\beta}$ from the \ess experiment for $2\%, 5\%$ and $10\%$ systematics. These values are obtained from the plots displayed in \fig \ref{fig:chisq}.}
\label{tab:bound_table}
\end{table}

Comparing the \ess results with other expected limits from upcoming LBL experiments in \eq (\ref{eq:dune_bounds}), we find that assuming nominal conditions ($5\%$ systematics), \ess bounds are less stringent than the DUNE ones by about a factor of 2. This is due to the higher energy and longer baseline for DUNE, so the effect of LRFs is more pronounced.  However, ESSnuSB outperforms T2HK by approximately $20\%$. As mentioned earlier, systematic uncertainties play a noticeable role in placing bounds on the LRF potential by ESSnuSB, i.e., achieving a $2\%$ normalization uncertainty could improve the \vem and $V_{e\tau}$ constraints, making them comparable to future DUNE bounds. Overall, ESSnuSB is projected to set bounds on LRF parameters that are competitive with those from future LBL experiments such as DUNE and T2HK. Importantly, the complementarity of constraints from various neutrino sources, including accelerator, atmospheric, and solar neutrino data, provides a unique opportunity to significantly narrow the allowed parameter space for LRFs. By combining these results, the interplay between different datasets may uncover synergies that enhance sensitivity to LRF parameters and help elucidate the underlying physics of these new interactions.

\begin{figure}
\centering
\includegraphics[width=14.5cm,height=14.0cm]{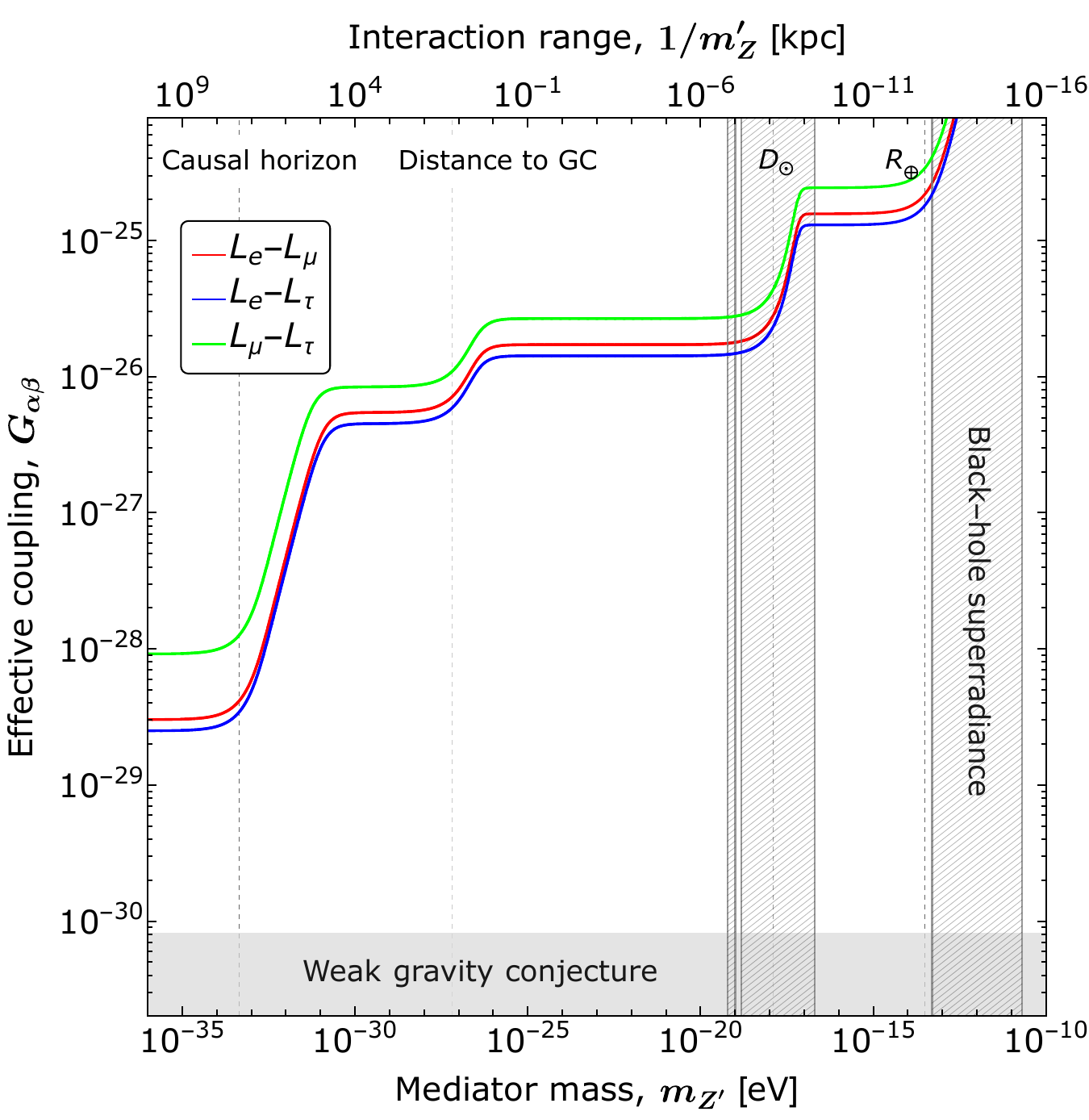}
\caption{\footnotesize \textit{\textbf{Sensitivity of \ess to exclude the parameter space of the LRF in the $m_{Z^\prime}-G_{\alpha\beta}$ plane.}} These exclusion regions are computed at $90\%$ C.L. by fixing the LRF potentials to their $90\%$ C.L. values presented in Table \ref{tab:bound_table}. See the text for more details on how the calculations were performed.}
        \label{fig:coupling_mass}
\end{figure}

In addition to the LRF potentials, $V_{\alpha\beta}$, we also put constraints on the actual parameters of the new neutrino-matter interaction, namely, the mass of the new gauge boson $m_{Z^\prime}$ and the effective gauge couplings $G_{\alpha\beta}$. Following the methodology presented in refs.~\cite{Bustamante:2018mzu, Singh:2023nek, Giarnetti:2024mdt}, we use \eqs (\ref{eq:pot_ve}) and (\ref{equ:pot_vmt}) to derive the limits on $m_{Z^\prime}$ and $G_{\alpha\beta}$. In order to take into account all the matter content present in the Universe, we consider neutrinos from different sources ranging at distances up to $10^3$ Gp away from the Earth. This corresponds to the mediator mass, $m_{Z^\prime}$ in the range $10^{-10}-10^{-35}$ eV and the LRF potentials originating from all the matter content of the Universe can be rewritten in terms of the contributions from effective potentials relevant at different distances, i.e.,

\begin{align}\label{eq:vsb_universe}
V_{\alpha\beta} = (V_{\alpha\beta})_{\rm Earth} +(V_{\alpha\beta})_{\rm Moon}+(V_{\alpha\beta})_{\rm Sun}+(V_{\alpha\beta})_{\rm MW}+(V_{\alpha\beta})_{\rm Cosm}\,.
\end{align}
To find the electron and neutron numbers for the LRF potentials from the Earth, an average density of a continuous distribution is modeled for the Earth such that we get $(N_e)_{\rm Earth} = (N_n)_{\rm Earth}\sim4\times 10^{51}$. The Moon and the Sun are assumed to be point-like electron and neutron sources which correspond to the number of electrons and neutrons as given by $(N_e)_{\rm Moon} = (N_n)_{\rm Moon}\sim5\times 10^{49}$ and $(N_e)_{\rm Sun}\approx (N_n)_{\rm Sun}\sim10^{57}$~\cite{Singh:2023nek}. In case of the Milky Way, the total matter content can be assumed to be distributed in the form of a thin and a thick disk, a central bulge and a diffuse gas~\cite{Dehnen:1996fa,Miller:2013nza,Bustamante:2018mzu}, yielding $(N_e)_{\rm MW} = (N_n)_{\rm MW}\sim10^{67}$. For the cosmological matter content, we use $(N_e)_{\rm Cosm}\approx (N_n)_{\rm Cosm}\sim 10^{79}$ adopted from refs.~\cite{Bustamante:2018mzu,Singh:2023nek}. Utilizing these values of electron and neutron numbers and using \eqs (\ref{eq:pot_ve}) and (\ref{equ:pot_vmt}), the contributing terms of LRF potentials from all sources can be computed provided that the values of $m_{Z^\prime}$ and $G_{\alpha\beta}$ are known. To constrain $m_{Z^\prime}$ and $G_{\alpha\beta}$, we use the $90\%$ C.L. limits on $V_{\alpha\beta}$ obtained in Table \ref{tab:bound_table} and vary the free parameters. The results are presented in \fig \ref{fig:coupling_mass} where red, blue and green curves correspond to the $L_e - L_\mu$, $L_e - L_\tau$ and \lmt symmetries, respectively. We also show the interaction range $\propto1/m_{Z^\prime}$ on the upper axis of the plot. It is worthwhile to mention that some astrophysical and cosmological phenomena, such as black-hole superradiance~\cite{Baryakhtar:2017ngi} and weak gravity conjecture~\cite{Arkani-Hamed:2006emk} may also exclude some parameter space of the LRFs, providing the non-oscillation exclusion limits. These regions are displayed by the grey bands in \fig \ref{fig:coupling_mass}.  From this figure, one can observe that the most stringent limit comes from the location of the causal horizon, which contains the highest number of electrons and neutrons. Therefore, the LRF potentials experienced by neutrinos from this location will be the largest. 

\subsection{Correlations}
\label{sec:correlation}
\begin{figure}
\hspace*{-1.5cm}
\includegraphics[width=7.5cm,height=6.5cm]{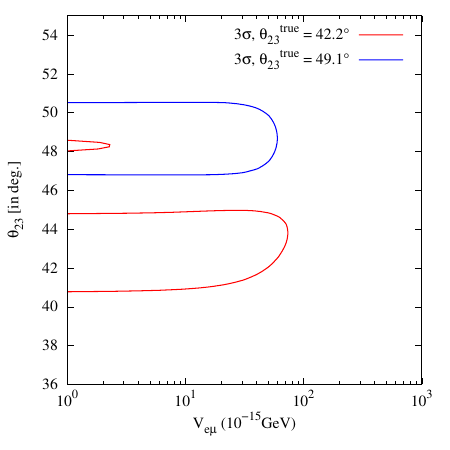}
\hspace*{0.5cm}
\includegraphics[width=7.5cm,height=6.5cm]{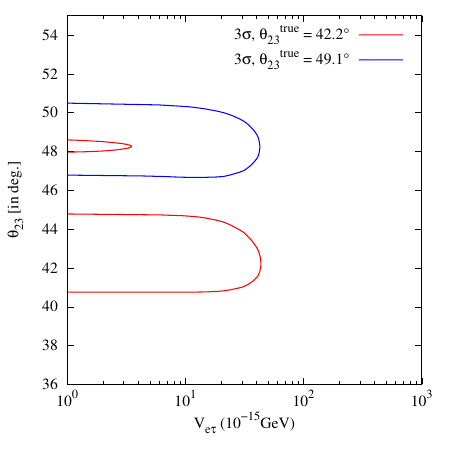}
\centering \includegraphics[width=8cm,height=6.5cm]{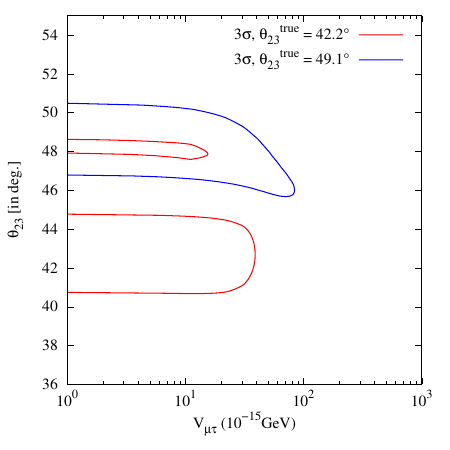}
\caption{\footnotesize \textbf{\textit{ Effect of long-range interactions in the determination of the  $\bold{\theta_{23}}$ octant.}} In all panels, two distinct true values for the mixing angle $\theta_{23}$ have been chosen, i.e., $\theta_{23}^{\rm true} = 42.2\degree$ and $49.1\degree$.}
        \label{fig:octant}
\end{figure}

In this subsection, we want to explore the correlations between the LRF potentials, \vab and the two poorly constrained standard neutrino oscillation parameters for ESSnuSB, namely $\delta_{\rm CP}$ and $\theta_{23}$. 
To conduct this analysis, the true event spectra were generated under the assumption of no LRFs, using the best-fit values for the standard oscillation parameters listed in Table \ref{tab:t1}. The fit was obtained by marginalizing over all standard oscillation parameters not explicitly shown, except for $\theta_{12}$ and $\Delta m^2_{21}$.
Figure \ref{fig:octant} displays the
3$\sigma$ allowed regions in the  $V_{\alpha\beta}-\theta_{23}$ plane. The upper-left and upper-right panels illustrate the correlations between   $V_{e\mu}~(\text{upper left}), V_{e\tau}~(\text{upper right})$ and $\theta_{23}$ whereas the lower panels represent the correlation in the $V_{\mu\tau}-\theta_{23}$ plane. We present the results for two different choices of true values of the mixing angle $\theta_{23}$, one in the lower octant ($42.2\degree$) and one in the upper octant ($49.1\degree$). These values correspond to the best-fits from Ref.~\cite{Esteban:2020cvm}, with and without the inclusion of the Super-Kamiokande atmospheric data. We see that, in the absence of LRFs, the \ess results suggest that the  $\theta_{23}$ octant degeneracy might not be resolved if the true value is $\theta_{23}=42.2\degree$. In this case, allowed values in the upper octant persist. However, in the presence of LRFs, this degeneracy appears to be resolved as the LRF potentials $V_{e\mu}$ and $V_{e\tau}$ tend to increase (see the upper left and right panels of \fig \ref{fig:octant}). A similar trend is observed for $V_{\mu\tau}$ (lower panel), although at relatively larger values of the potential. On the other hand, for $\theta_{23}=49.1\degree$ and $V_{\alpha\beta}\to0$, the octant degeneracy appears to be already broken. Even in the presence of LRFs, we do not see any octant ambiguity when the true value of $\theta_{23} = 49.1\degree$.  As discussed in the previous section, we remark that the octant degeneracy breaking is mainly due to the appearance channel. Indeed, if we only consider the disappearance channel, the octant ambiguity plays a major role in the analysis.



\begin{figure}[H]
\hspace*{-1.5cm}
\includegraphics[width=7.5cm,height=6.5cm]{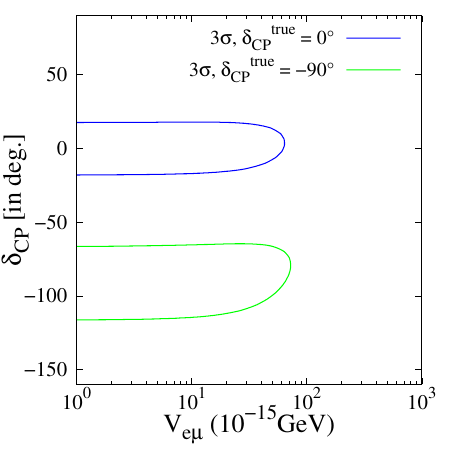}
\hspace*{0.5cm}
\includegraphics[width=7.5cm,height=6.5cm]{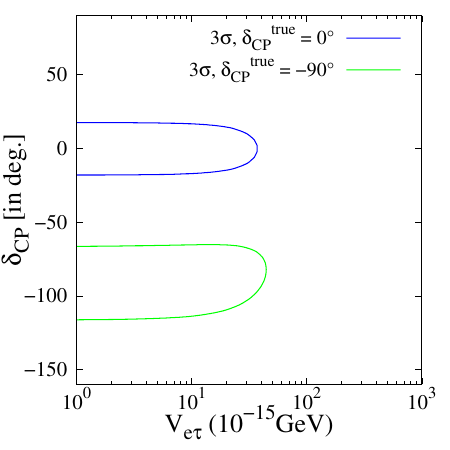}
\centering \includegraphics[width=8cm,height=6.5cm]{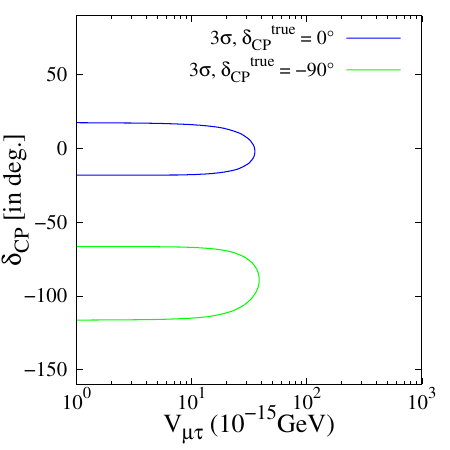}
\caption{\footnotesize \textbf{\textit{ Effect of long-range interactions on the determination of $\bold{\delta_{\rm CP}}$}}. In all the panels, two distinct true values for the leptonic CP-violating phase have been chosen, $\delta_{\rm CP}^{\rm true} = 0\degree, -90\degree$.}
        \label{fig:dcp_vab}
\end{figure}

In Fig. \ref{fig:dcp_vab}, we present the results in the \vem $-$ \dcp (upper-left panel), \vet $-$ \dcp (upper-right panel) and \vmt $-$ \dcp (lower panel) planes for two true values of \dcp corresponding to maximal CPV ($\delta_{\rm CP}=-90^\circ$) and no CPV ($\delta_{\rm CP}=0^\circ$). The marginalization scheme used in this analysis follows the same procedure as previously described, where all other oscillation parameters, except $\theta_{12}$ and $\Delta m^2_{21}$, are marginalized.  In this case, we observe no significant correlations between \dcp and any of the LRF potentials, $V_{e\mu}$, \vet and $V_{\mu\tau}$. However, it is worth noting that the effects of \vab on the determination of 
\dcp could become significant if the LRF strengths are large enough to achieve the sensitivity of the ESSnuSB experiment. In such a scenario, those values might become measurable, introducing a potential influence on \dcp determination. We will delve deeper into this possibility and its implications in the next section.

\section{CPV sensitivity of \ess in the presence of LRFs}
\label{sec:CPV_sensitivity}

\begin{figure}[H]
\centering
\includegraphics[width=15.5cm,height=10.0cm]{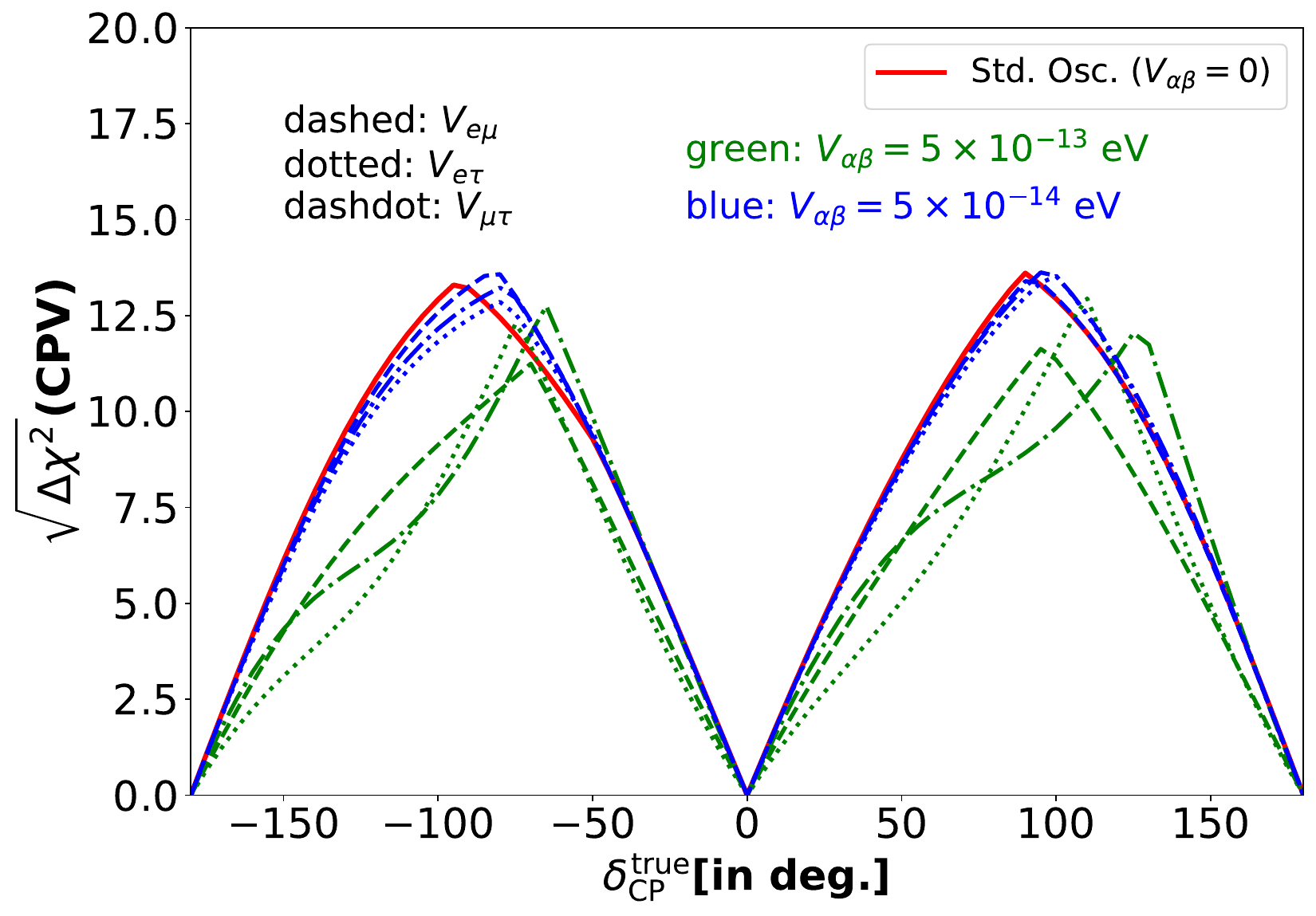}
\caption{\footnotesize \textit{\textbf{CP-violation sensitivity of ESSnuSB for different values of LRF potentials \vab and symmetries.}} Here, both true and test hypotheses assume the presence of LRFs. Standard oscillation (\vab = 0) is shown by the solid red curve.}
        \label{fig:cpv1}
\end{figure}

In this section, we examine how the LRF potentials influence the CP-violation sensitivity of the ESSnuSB experiment. This analysis is crucial, as the primary aim of ESSnuSB is to achieve precise measurements of $\delta_{\rm CP}$. It is worth noting that, in the case of maximal CP violation (\dcp $=\pm90\degree$),  the sensitivity of \ess can reach up to 12.5$\sigma$ and it can also achieve at least 5$\sigma$ sensitivity for approximately 75\% of the other possible values of \dcp\cite{Alekou:2022emd,ESSnuSB:2023lbg}. This surpasses the sensitivity of all upcoming next-generation LBL neutrino oscillation experiments~\cite{Agarwalla:2022xdo}. It is, therefore, vital to determine whether the presence of new physics, such as long-range interactions of neutrinos with matter, could jeopardize this capability or not.  To do this, we generate the true event spectrum by varying \dcp (true) over the full range $[-180\degree,180\degree]$ and compare this with \dcp $=0\degree$ or $180\degree$ in the test. The same value of LRF potentials \vab is considered in both true and test event spectra.  The CPV-sensitivity plots are displayed in \fig \ref{fig:cpv1} in units of $\sqrt{\Delta \chi^2}$, where
\begin{equation}
    \Delta\chi^2=\chi^2(V_{\alpha\beta},\mathrm{CPV})-\chi^2(V_{\alpha\beta},\delta_\mathrm{CP}=0\degree,180^\circ) \, .
\end{equation}

\begin{figure}[H]
\centering
\includegraphics[width=15.5cm,height=10.0cm]{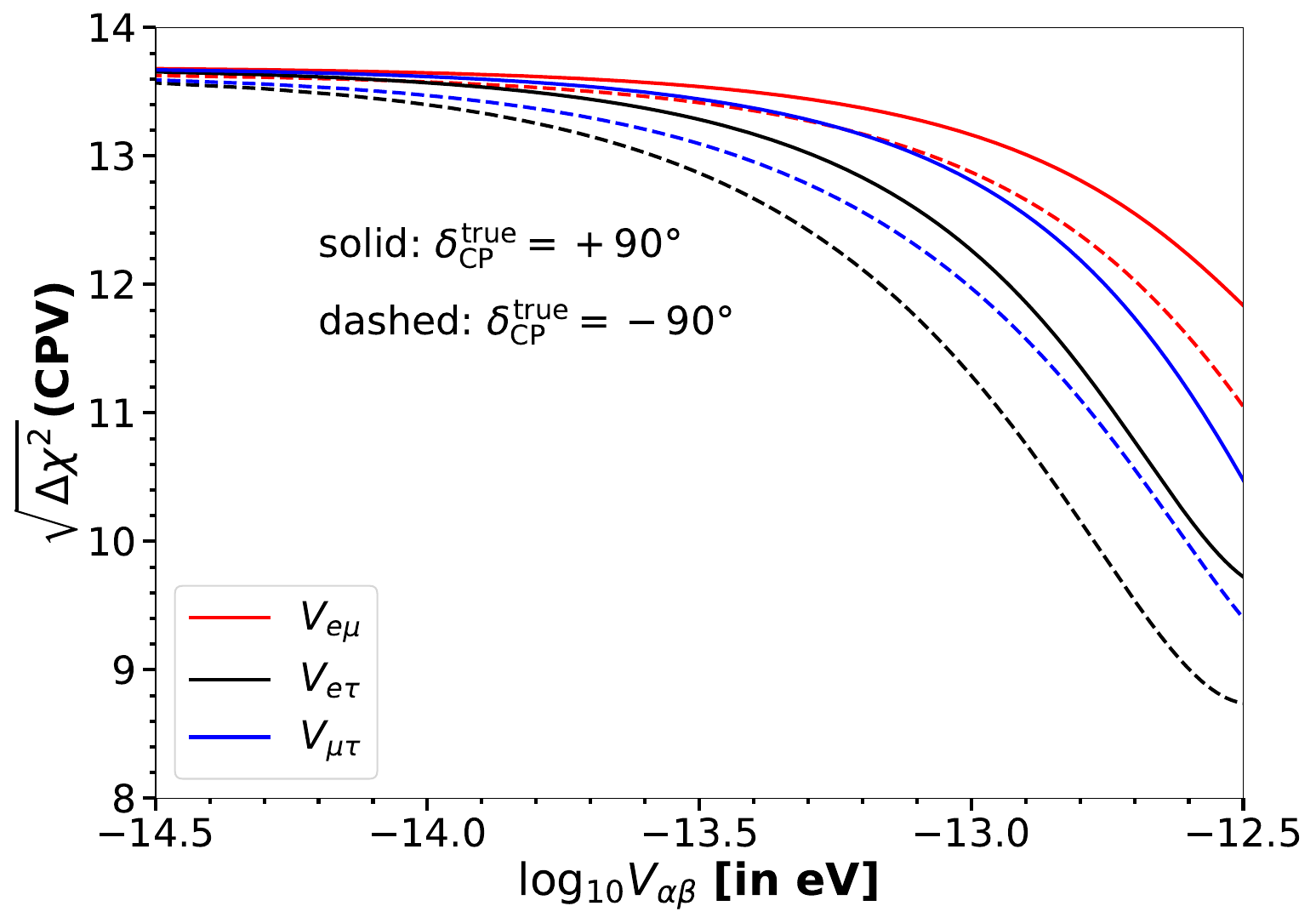}
\caption{\footnotesize \textit{\textbf{CP-violation sensitivity of ESSnuSB as a function of long-range potential $V_{\alpha\beta}$.}} The solid (dashed) curves correspond to the true value of \dcp $=+90 \degree(-90\degree)$. }
        \label{fig:cpv2}
\end{figure}

In this figure, the red curve represents the sensitivity in the standard oscillation scenario (\vab $=0)$. The dashed, dotted and dashdot curves are for the potentials $V_{e\mu}$, \vet and $V_{\mu\tau}$, respectively. The sensitivity curves plotted in blue colour correspond to the value of \vab = $5\times10^{-14}$ eV, which is comparable to the \ess constraints, while the curves in green are computed for the LRF potentials, \vab = $5\times10^{-13}$ eV, a much larger potential value than the \ess bounds.  We notice that for small values of the new potentials, the \ess CP-violation sensitivity remains intact with some negligible impact on its sensitivity around \dcp $=\pm 90\degree$. However, for large values of the LRF potentials, the $\Delta \chi^2$ changes and the positions of the sensitivity maxima are also slightly shifted. To understand this in more detail, we compute the CPV sensitivity as a function of the LRF potentials, $V_{\alpha\beta}$. The results are displayed in \fig \ref{fig:cpv2} for two choices of $\delta_{\rm CP}$, i.e.~$+90\degree$ (solid curves) and $-90\degree$ (dashed curves). We can observe that, for small values of \vab ($ \lesssim 10^{-14}$ eV), the \dcp sensitivity of \ess more or less does not change for all three cases of $V_{\alpha\beta}$. However, with increasing $V_{\alpha\beta}$, the sensitivity decreases, especially for the potential corresponding to the \letau symmetry. The reason is that when \vab are small, they appear as a correction to the standard probability and mildly affect the \dcp sensitivity, whereas, for large values of $V_{\alpha\beta}$, new resonances might appear, causing a significant drop in the CPV sensitivity of ESSnuSB.

\section{CP precision of \ess in the presence of LRFs}\label{sec:cpv_precision}

In this section, we will try to understand the impact of LRF potentials on the uncertainty of \dcp measurement by the \ess experiment. Since the primary objective of \ess is to perform a precision measurement of $\delta_{\rm CP}$ in addition to discovering it (if next-generation LBL experiments fall short), it is imperative to see how new physics affects this capability of ESSnuSB. In Ref.~\cite{Alekou:2022emd,ESSnuSB:2023lbg}, it has been shown that the optimal baseline of 360 km allows the \ess experiment to measure $\delta_{\rm CP}$ with a 1$\sigma$ uncertainty of less than $7.5\degree$ for all possible values of $\delta_{\rm CP}$. Remarkably, the experiment achieves its best precision, $\Delta\delta_\mathrm{CP} = 5^\circ$, for CP-conserving values. Such a level of accuracy is unparalleled, as it surpasses the capabilities of next-generation LBL experiments, emphasizing the transformative potential of ESSnuSB in this area of research. 

\begin{figure}[H]
\hspace*{-1.5cm}
\includegraphics[width=8.8cm,height=7.5cm]{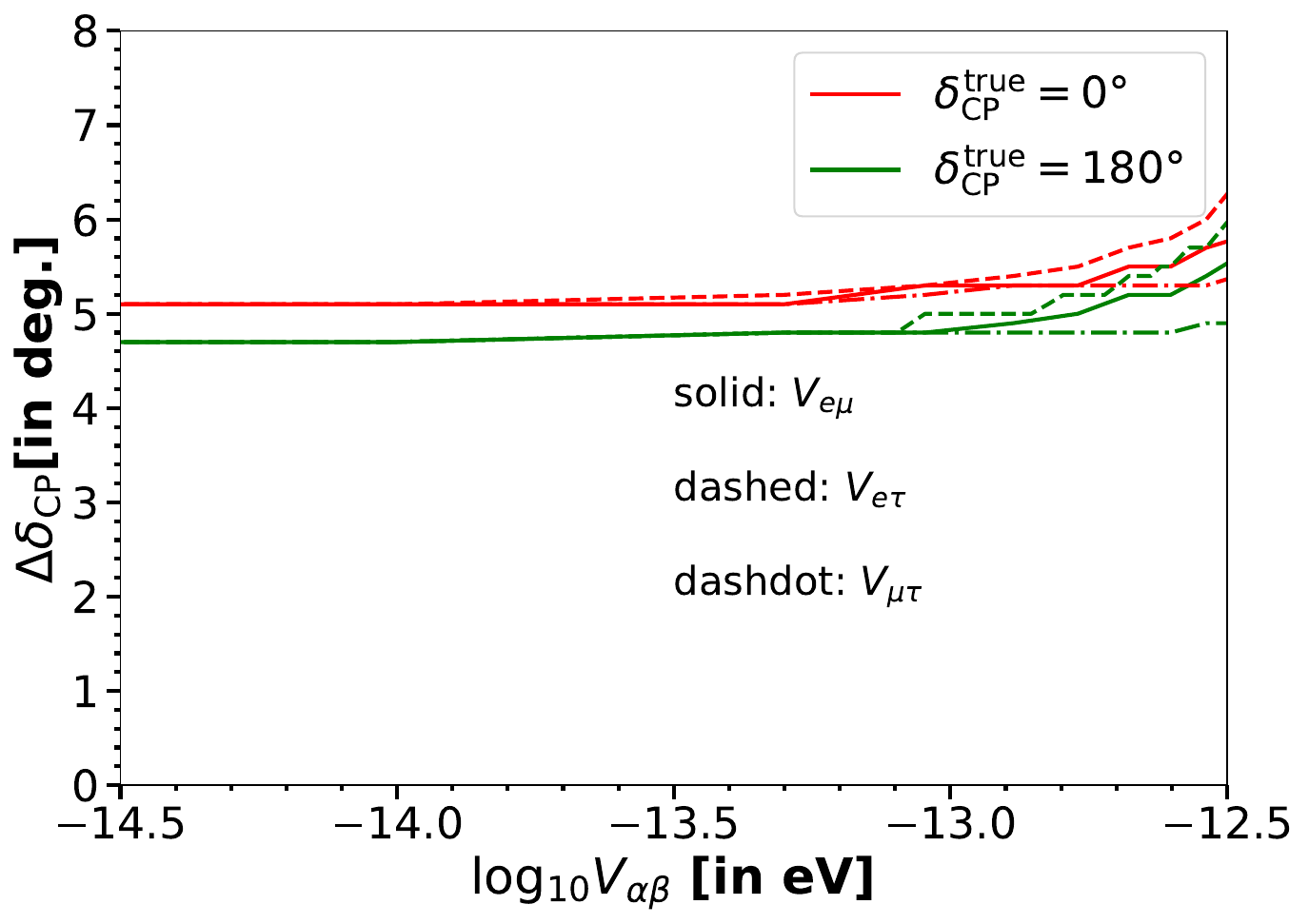}
\hspace*{-0.01cm}
\includegraphics[width=8.8cm,height=7.5cm]{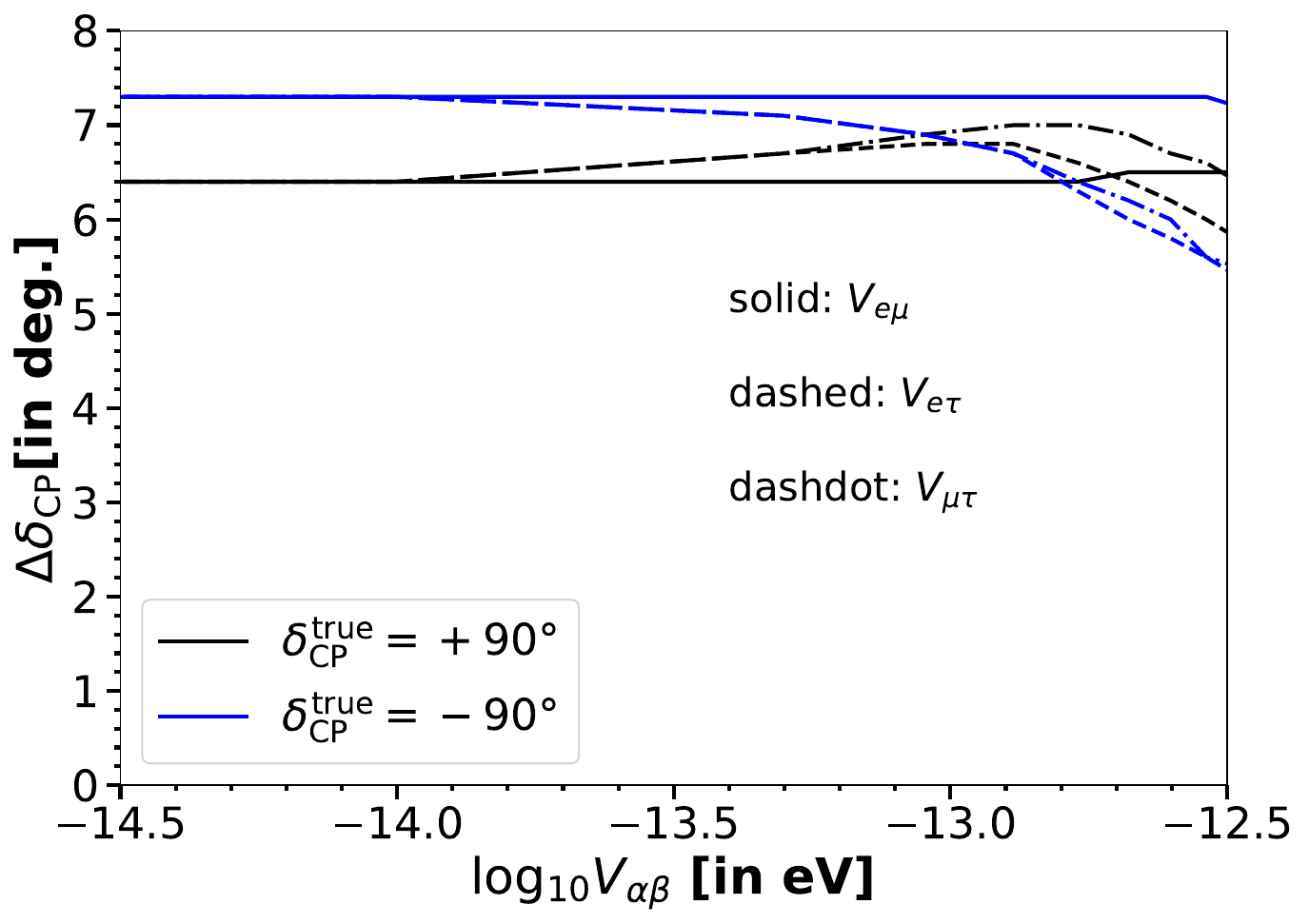}
\caption{\footnotesize\textit{\textbf{$1\sigma$ precision on the measurement of $\delta_{\rm CP}$ at ESSnuSB as a function of the LRF potential $V_{\alpha\beta}$, for three
different choices of symmetries.}} The LRF potential $V_{\alpha\beta}$ is present in both true and test data. The left (right) plot corresponds
to the true values of $\delta_{\rm CP} = 0\degree$ and $180\degree$ ($\pm 90\degree$).}
        \label{fig:cpv_precision}
\end{figure}

In Fig.~\ref{fig:cpv_precision}, we present the projected 1$\sigma$ uncertainty in the measurement of $\delta_{\rm CP}$ for two scenarios: the CP-conserving values ($\delta_{\rm CP} = 0\degree$ and $180\degree$, shown in the left panel) and the maximally CP-violating values ($\delta_{\rm CP} = \pm90\degree$, shown in the right panel), as functions of the LRF potentials, $V_{\alpha\beta}$. In both panels, solid, dashed and dashdot curves represent LRF potentials, namely, $V_{e\mu}$, \vet and  $V_{\mu\tau}$, respectively. From \fig \ref{fig:cpv_precision} it is evident that the effects of LRF potentials on the \dcp precision of \ess are negligible. Even when the values of all three potentials are large enough (almost an order of magnitude larger than the \ess bounds), the effects of \vab are not significant enough to meaningfully degrade the performance of ESSnuSB. Specifically, for the maximally CP-violating values ($\delta_{\rm CP} = \pm90\degree$), illustrated in the right panel of \fig \ref{fig:cpv_precision}, the experiment can achieve a robust precision of $\Delta\delta_{\rm CP} < 7.5\degree$, as long as \vab remain below $2 \times 10^{-14}$ eV. For the CP-conserving values ($\delta_{\rm CP} = 0\degree$ and $180\degree$), illustrated in the left panel of \fig \ref{fig:cpv_precision}, the precision is even better, with $\Delta\delta_{\rm CP} \lesssim 7\degree$ across the entire range of LRF potentials, $V_{\alpha\beta}$.

These results highlight the resilience of \ess in maintaining high precision in $\delta_{\rm CP}$ measurements, even in the presence of LRFs, further demonstrating its capability to probe CP violation with unprecedented accuracy.

\section{Summary and conclusions}
\label{sec:summaery}
In this paper, we explored the capabilities of the \ess experiment to set bounds on the effects of LRFs in neutrino oscillations. In the presence of additional $U(1)$ gauge symmetries in the particle physics Lagrangian, a new vector mediator $Z^\prime$ might be responsible for new interactions between SM particles. In the case of a very light mediator, such interactions might occur at very long distances and feebly interacting particles like neutrinos could provide valuable information about them. For instance, in neutrino oscillation experiments, LRFs modify matter effects in the neutrino oscillation probabilities introducing new terms in the Hamiltonian.  We considered three different $U(1)$ symmetries, namely $L_e-L_\mu$, $L_e-L_\tau$ and $L_\mu-L_\tau$. We demonstrated how the ESSnuSB setup could provide a good environment to search for LRFs. In particular, using nominal conditions (5\% systematics), we observed that \ess could be able to set 90\% C.L. limits on $V_{e\mu}<2.99\times10^{-14}$ eV, $V_{e\tau}<2.05\times10^{-14}$ eV and $V_{\mu\tau}<1.81 \times 10^{-14}$ eV. The bounds on such parameters have been obtained
by means of a standard $\chi^2$ analysis performed using the GloBES software. Among the upcoming next-generation LBL experiments, only DUNE is expected to outperform ESSnuSB, while T2HK will set weaker limits \cite{Singh:2023nek,Giarnetti:2024mdt}. The \ess bounds might become comparable to the DUNE ones if systematic uncertainties in both the appearance and disappearance channels are reduced for the \ess experiment. We explored the correlations between the LRF parameters and the most unknown oscillation parameters, namely $\theta_{23}$ and $\delta_{\rm CP}$. We found that the octant degeneracy of $\theta_{23}$ is broken in the presence of LRFs when $\theta_{23}^{\rm true} = 42.2\degree$. We also could not observe any strong correlation between \dcp and the LRF potentials $V_{\alpha\beta}$.

Finally, we addressed another crucial point in the context of the \ess experiment: the robustness of its most important measurement, namely the $\delta_{\rm CP}$ determination. We observed that, even in the presence of LRFs, both the CPV sensitivity and the \dcp precision remain unaltered except in the case of extremely large LRF potentials ($V_{\alpha\beta}\gg10^{-13}$).\\

\section*{Acknowledgements}

Funded by the European Union. Views and opinions expressed are however those of the author(s) only and do not necessarily reflect those of the European Union. Neither the European Union nor the granting authority can be held responsible for them.

We acknowledge further support provided by the following research funding agencies: Centre National de la Recherche Scientifique, France; Deutsche Forschungsgemeinschaft, Germany, Projektnummer 423761110; Ministry of Science and Education of Republic of Croatia grant No. KK.01.1.1.01.0001; the European Union’s Horizon 2020 research and innovation programme under the Marie Skłodowska-Curie grant agreement No 860881-HIDDeN; the European Union NextGenerationEU, through the National Recovery and Resilience Plan of the Republic of Bulgaria, project No. BG-RRP-2.004-0008-C01; as well as support provided by the universities and laboratories to which the authors of this report are affiliated, see the author list on the first page. 

\newpage
\appendix

\section{LRF induced by other $U(1)$ symmetries}\label{sec:appendix}

\begin{figure} 
\hspace*{-1cm}
     \centering
     \begin{subfigure}[b]{0.5\textwidth}
         \centering
         \includegraphics[width=\textwidth, height = 6cm]{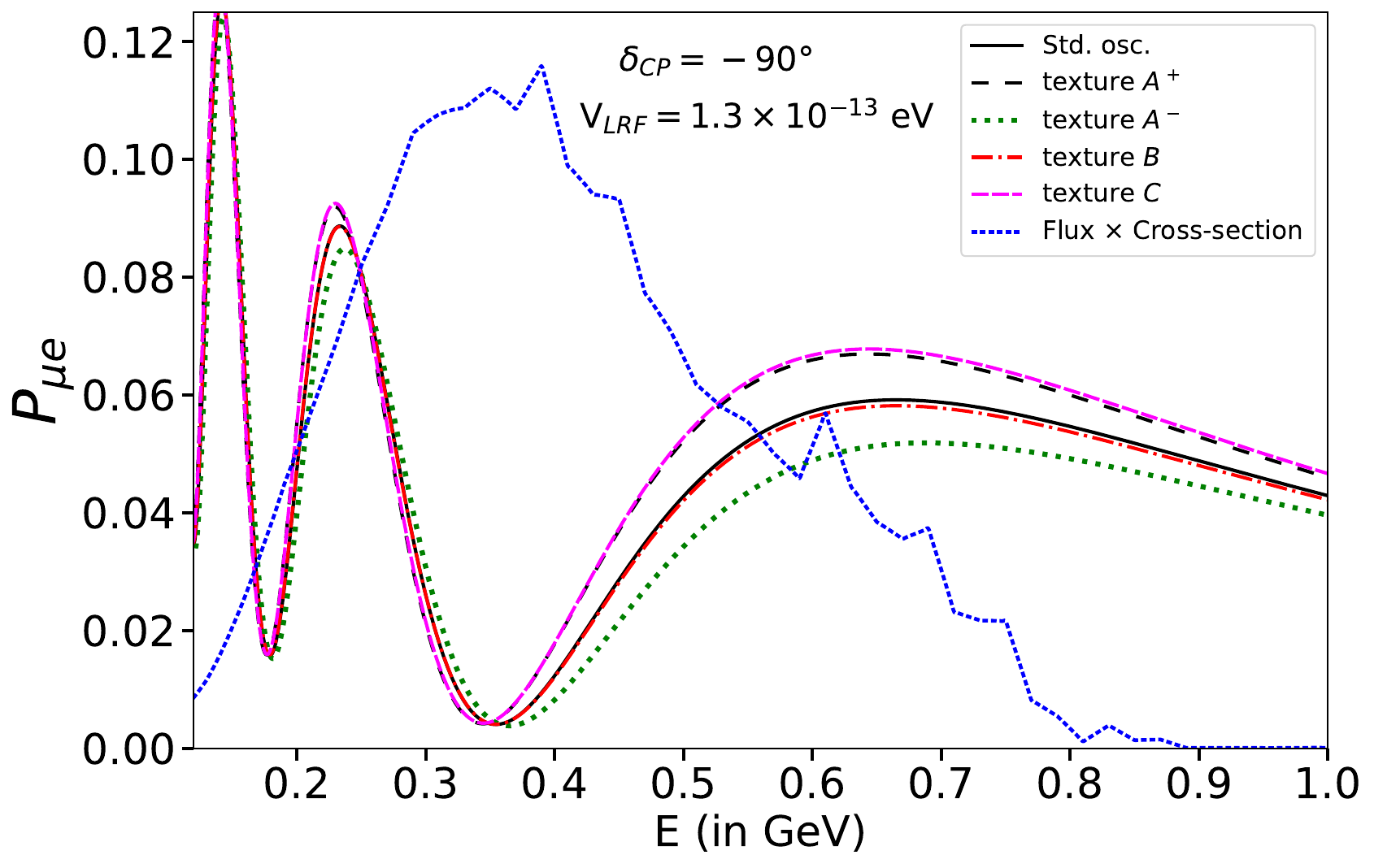}
         \caption{}
         
     \end{subfigure}
     \hfill
     \begin{subfigure}[b]{0.5\textwidth}
     \centering
     \includegraphics[width=\textwidth,height = 6cm]{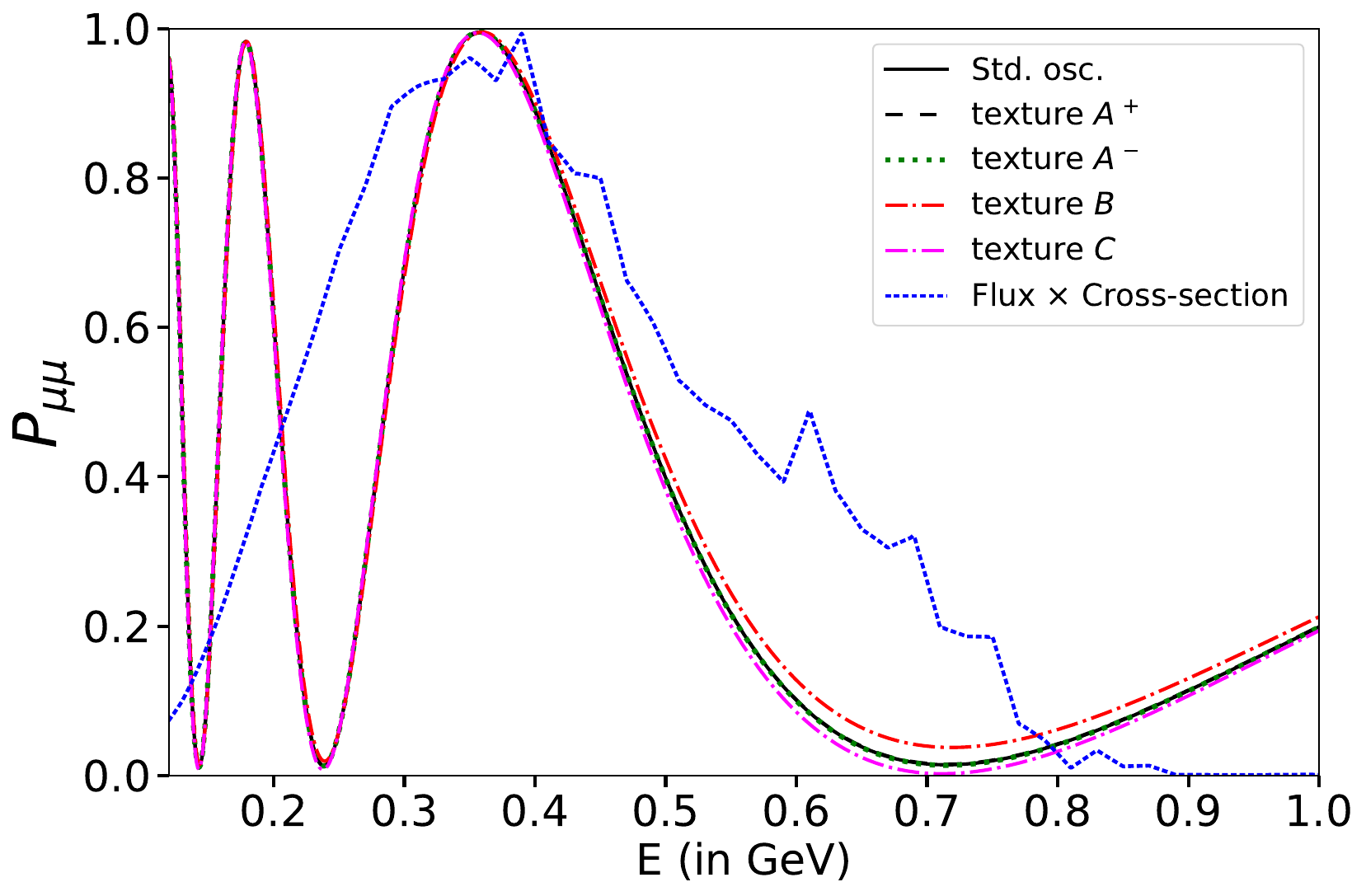}
     \caption{}
    
     \end{subfigure}
     \hfill
     \hspace*{-1cm}
    \begin{subfigure}[b]{0.5\textwidth}
         \centering
         \includegraphics[width=\textwidth,height = 6cm]{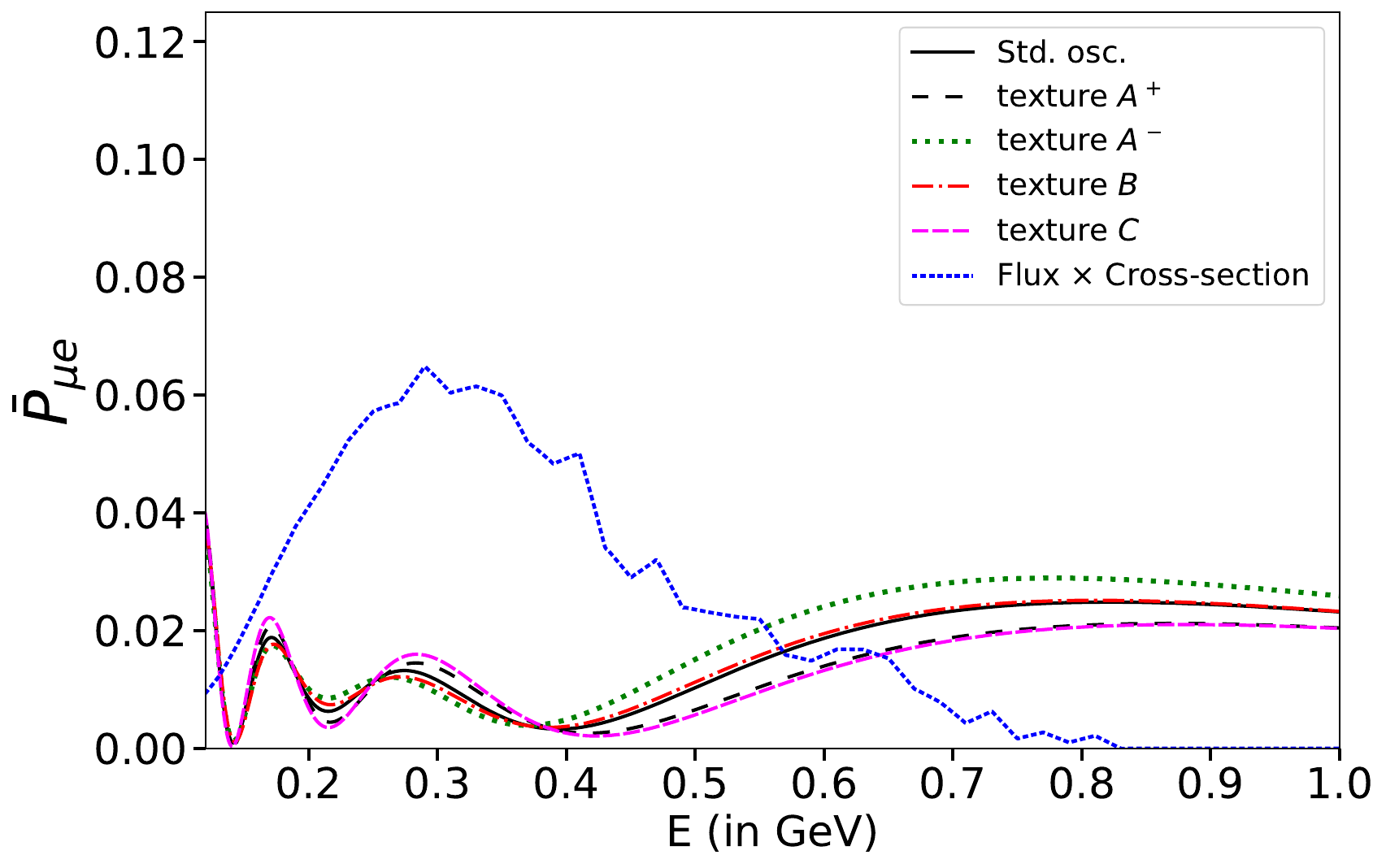}
         \caption{}
     \end{subfigure}
     \hfill
     \begin{subfigure}[b]{0.5\textwidth}
         \centering
         \includegraphics[width=\textwidth,height = 6cm]{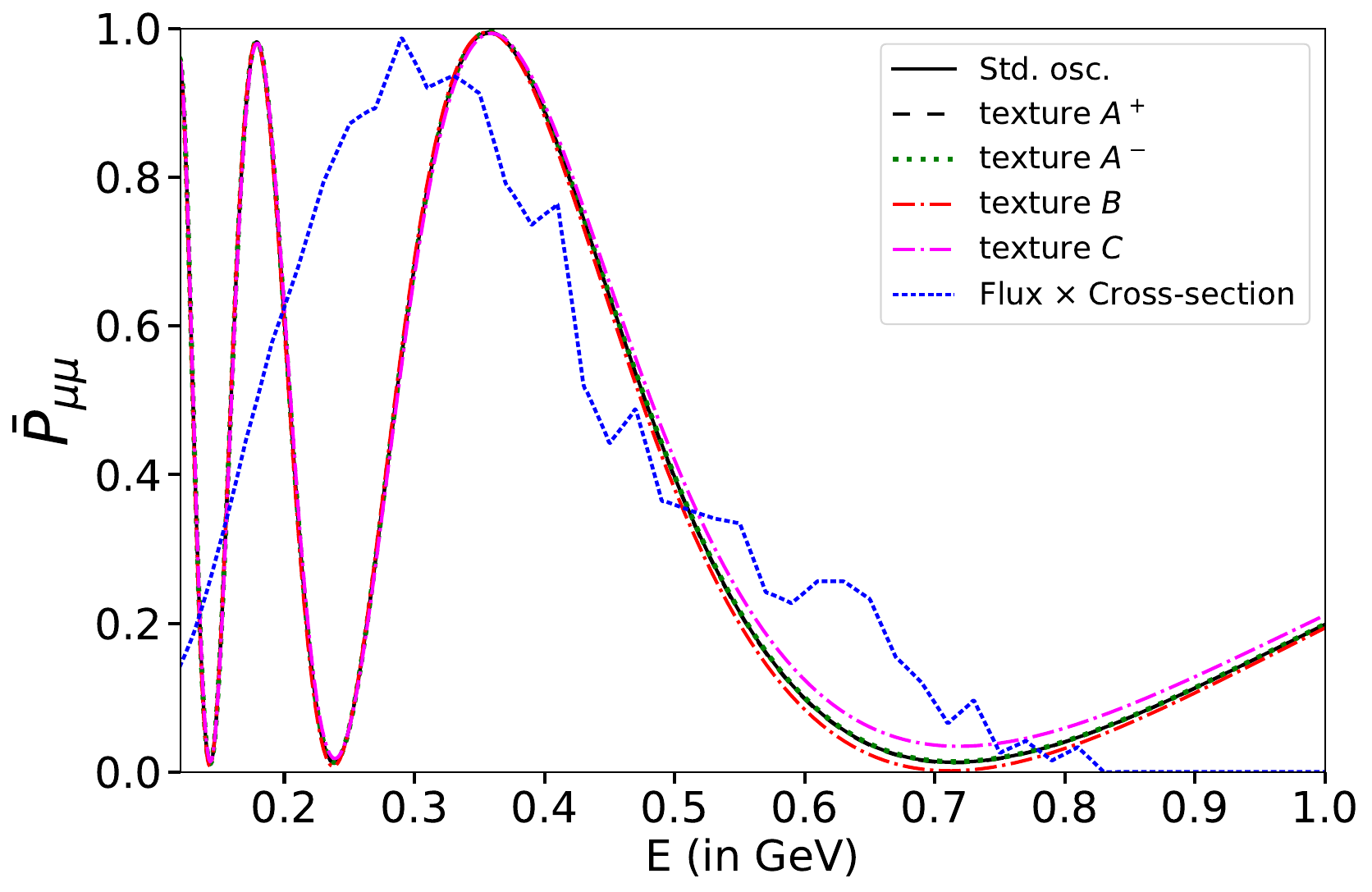}
         \caption{}
     \end{subfigure}
     \hfill
     \centering
     \caption{\footnotesize\textbf{\textit{Appearance (left panel) and disappearance (right panel) neutrino (top panel) and antineutrino (bottom panel) oscillation probabilities as functions of neutrino energy in the presence of LRF potentials induced by other sorts of anomaly-free symmetries.}}}
        \label{fig:prob_plethora}
\end{figure}

In Ref. \cite{Agarwalla:2024ylc}, other possibilities for the $U(1)$ symmetries, which might generate new interactions modifying the neutrino oscillation probabilities, have been explored for the first time. In the context of neutrino oscillation, the different anomaly-free combinations of baryon number $B$ and lepton numbers $L_\alpha$ \cite{Coloma:2020gfv,Davoudiasl:2011sz,Araki:2012ip,delaVega:2021wpx,Farzan:2016wym,Joshipura:2019qxz,Almumin:2022rml} can only modify the diagonal entries of the matter potential part of the oscillation Hamiltonian (see Table~\ref{tab:plethora_bound} for a list of the symmetries and Ref.~\cite{Agarwalla:2024ylc} for details). Thus, regardless of the specific combination of the charges of the particles under the specific symmetries, the only textures of the LRF matrix in the Hamiltonian not discussed in our previous analyses are \cite{Agarwalla:2024ylc}
\begin{equation}
 \label{equ:lrf_matrix_onlyone}
 {V}_{\alpha\beta}
 =
 \left\{
  \begin{array}{ll}
   {\rm diag}(\pm V_{LRF}, 0, 0), & (\rm{textures} \quad A^{\pm})  \\
   {\rm diag}(0, -V_{LRF}, 0), & (\rm{texture} \quad B)\\
   {\rm diag}(0,0, -V_{LRF}), & (\rm{texture} \quad C)\\   
  \end{array}
 \right. \;.
\end{equation}
To quantitatively discuss the effects of the LRF mediated by the symmetries generating the $V_{\alpha\beta}$ matrices in the oscillation Hamiltonian, we show in Fig. \ref{fig:prob_plethora} the $\overset{(-)}{\nu_\mu}\to\overset{(-)}{\nu_e}$ (panels a and c) and $\overset{(-)}{\nu_\mu}\to\overset{(-)}{\nu_\mu}$ (panels b and d) oscillation probabilities in the energy range interesting for ESSnuSB. In the appearance probability, the textures $A^+$ and $C$ have the same effect, enhancing (reducing) the neutrino (antineutrino) probability at oscillation maximum by roughly the same amount. The $A^-$ texture, on the other hand, has approximately the same but opposite effect of the $A^+$ texture. The $B$ texture does not significantly affect the oscillations. This behaviour also explains why the $L_e-L_\tau$ symmetry modifies the probabilities more than $L_e-L_\mu$ and $L_\mu-L_\tau$; indeed in this specific case, the effects of the $A^+$ and $C$ texture are summed, enhancing the probabilities more.
The disappearance probabilities, on the other hand, are almost unaffected by the $A^\pm$ textures and for the neutrino case, $P_{\mu\mu}$ is enhanced (reduced) at the minimum by the $B$ ($C$) textures. The antineutrino disappearance probability exhibits opposite behaviour compared to the neutrino case when textures $B$ and $C$ are considered. 

\begin{figure}
     \centering
\includegraphics[width=11.8cm,height=9.0cm]{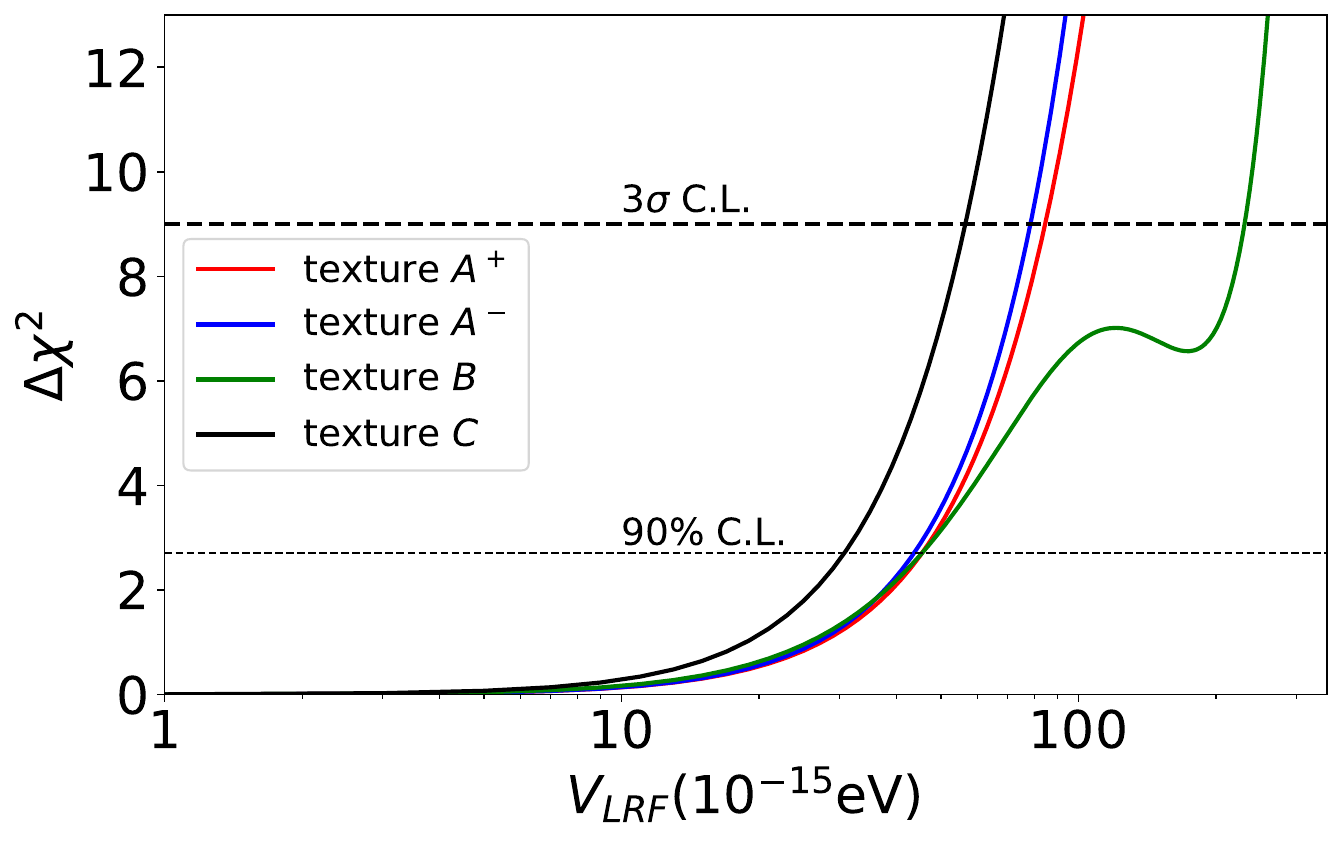}
\caption{\footnotesize\textbf{\textit{Sensitivity of ESSnuSB to constrain the LRF potentials induced by different $U^\prime$ symmetries mentioned in Ref.~\cite{Agarwalla:2024ylc}.}} In this case, we have used the standard $5\%$ systematics of the \ess experiment.}
        \label{fig:bounds_plethora}
\end{figure}

We finally study the sensitivity of the ESSnuSB experiment to the $V_{LRF}$ parameters in the four studied cases in Fig.~\ref{fig:bounds_plethora}. We summarize in Table \ref{tab:plethora_bound} the $3\sigma$ bounds for the four textures obtained using the same procedure described in Sec. \ref{sec:bounds} with 5\% systematics. The bounds in these cases are, in general, less stringent than in the $L_\alpha-L_\beta$ symmetries cases. In particular, the texture $C$ gives the tightest bound on $V_{LRF}$, while the texture $B$ is the looser. It is interesting to notice that since texture $B$ does not affect significantly the appearance channel but only the disappearance one, the octant degeneracy causes a reduction of the $3\sigma$ sensitivity for $V_{LRF}$.


\begin{table}[H]
    \centering
    \begin{tabular}{|c|c|c|c|}
    \hline
    \multirow{2}{*}{LRF Potential Textures} & 
    \makecell{$3\sigma$ C.L. \\ $(\times 10^{-14})$ eV} & 
    \makecell{$90\%$ C.L. \\ $(\times 10^{-14})$ eV} & 
    \makecell{$U(1)$ symmetries from Ref.~\cite{Agarwalla:2024ylc}} \\
    \hline
    $A^+$ & 8.45 & 4.55 &
    \begin{tabular}{c}
        $B-3L_e$ \\
        $L-3L_e$ \\
        $B_y - \frac{3}{2} (L_\mu - L_\tau)$ \\
        $L_e - \frac{1}{2} (L_\mu - L_\tau)$
    \end{tabular} \\
    \hline
    $A^-$ & 7.85 & 4.37 &
    \begin{tabular}{c}
        $L_e + 2L_\mu + 2L_\tau$ \\
        $B + L_\mu + L_\tau$
    \end{tabular} \\
    \hline
    $B$ & 23.09 & 4.55 &
    \begin{tabular}{c}
        $B - 3L_\mu$ \\
        $L - 3L_\mu$
    \end{tabular} \\
    \hline
    $C$ & 5.65 & 3.08 &
    \begin{tabular}{c}
        $B - 3L_\tau$ \\
        $L - 3L_\tau$
    \end{tabular} \\
    \hline
    \end{tabular}
     \caption{\footnotesize\textbf{\textit{Constraints on LRF potential $V_{\alpha\beta}$, considering other $U(1)$ symmetries, using the \ess experiment for $5\%$ systematics.}} These values are obtained from the plots displayed in \fig \ref{fig:bounds_plethora}. In the last column, we show the symmetries mentioned in Ref.~\cite{Agarwalla:2024ylc}, where $L$ is the lepton number, $B$ is the baryon number and $B_y=B_1-yB_2-(3-y)B_3$ with $B_i$ being the baryon numbers of the quark families and $y$ an arbitrary constant.}
\label{tab:plethora_bound}
\end{table}




\newpage
\bibliographystyle{JHEP}
\bibliography{reference.bib,decoherence_ref}

\end{document}

%% file: authors_copy.tex
\newcommand{\authorlist}{
\author[1]{J.~Aguilar,}
\author[2]{M.~Anastasopoulos,}
\author[3]{D.~Barčot,}
\author[4]{E.~Baussan,}
\author[2]{A.K.~Bhattacharyya,}
\author[2]{A.~Bignami,}
\author[5,6]{M.~Blennow,}
\author[7]{M.~Bogomilov,}
\author[2]{B.~Bolling,}
\author[4]{E.~Bouquerel,}
\author[8]{F.~Bramati,}
\author[8]{A.~Branca,}
\author[8]{G.~Brunetti,}
\author[1]{I.~Bustinduy,}
\author[9]{C.J.~Carlile,}
\author[9]{J.~Cederkall,}
\author[10]{T.~W.~Choi,}
\author[5,6]{S.~Choubey,}
\author[9]{P.~Christiansen,}
\author[2,11]{M.~Collins,}
\author[8]{E.~Cristaldo Morales,}
\author[12]{P.~Cupia\l,}
\author[13]{D.~D'Ago,}
\author[2]{H.~Danared,}
\author[4]{J.~P.~A.~M.~de~Andr\'{e},}
\author[4]{M.~Dracos,}
\author[14]{I.~Efthymiopoulos,}
\author[10]{T.~Ekel\"{o}f,}
\author[2]{M.~Eshraqi,}
\author[15]{G.~Fanourakis,}
\author[16]{A.~Farricker,}
\author[17,18]{E.~Fasoula,}
\author[19]{T.~Fukuda,}
\author[2]{N.~Gazis,}
\author[15]{Th.~Geralis,}
\author[3,*]{M.~Ghosh,\orcidlink{0000-0003-3540-6548},\note[*]{Corresponding authors}}
\author[20,*]{A.~Giarnetti,\orcidlink{0000-0001-8487-8045},}
\author[21,22]{G.~Gokbulut,}
\author[23,24,*]{A.~Gupta\orcidlink{0000-0002-7247-2424},}
\author[25]{C.~Hagner,}
\author[3]{L.~Halić,}
\author[22]{M.~Hooft,}
\author[9]{K.~E.~Iversen,}
\author[22]{N.~Jachowicz,}
\author[3]{M.~Jakkapu,}
\author[2]{M.~Jenssen,}
\author[2]{R.~Johansson,}
\author[17]{E.~Kasimi,}
\author[21]{A.~Kayis Topaksu,}
\author[2]{B.~Kildetoft,}
\author[3]{B.~Kliček,}
\author[17,18]{K.~Kordas,}
\author[3]{B.~Kovač,}
\author[26]{A.~Leisos,}
\author[2,9,**]{M.~Lindroos, \note[**]{deceased}}
\author[27]{A.~Longhin,}
\author[2]{C.~Maiano,}
\author[28]{D.~Majumdar,}
\author[8]{S.~Marangoni,}
\author[20]{S.~Marciano,}
\author[22]{J.~G.~Marcos,}
\author[14]{C.~Marrelli,}
\author[20,*]{D.~Meloni,\orcidlink{0000-0001-7680-6957},}
\author[13]{M.~Mezzetto,}
\author[2]{N.~Milas,}
\author[1]{J.L.~Muñoz,}
\author[22]{K.~Niewczas,}
\author[21]{M.~Oglakci,}
\author[5,6]{T.~Ohlsson,}
\author[10]{M.~Olveg\r{a}rd,}
\author[24]{M.~Pari,}
\author[2]{D.~Patrzalek,}
\author[7]{G.~Petkov,}
\author[17,18]{Ch.~Petridou,}
\author[4]{P.~Poussot,}
\author[15]{A~Psallidas,}
\author[13]{F.~Pupilli,}
\author[29]{D.~Saiang,}
\author[17,18]{D.~Sampsonidis,}
\author[8]{A.~Scanu,}
\author[4]{C.~Schwab,}
\author[1]{F.~Sordo,}
\author[15]{G.~Stavropoulos,}
\author[3]{M.~Stipčević,}
\author[2]{R.~Tarkeshian,}
\author[8]{F.~Terranova,}
\author[25]{T.~Tolba,}
\author[2]{E.~Trachanas,}
\author[7]{R.~Tsenov,}
\author[26]{A.~Tsirigotis,}
\author[17]{S.~E.~Tzamarias,}
\author[22]{M.~Vanderpoorten,}
\author[7]{G.~Vankova-Kirilova,}
\author[30]{N.~Vassilopoulos,}
\author[5,6]{S.~Vihonen,}
\author[4]{J.~Wurtz,}
\author[4]{V.~Zeter,}
\author[15]{O.~Zormpa}
}

\affiliation[1]{Consorcio ESS-bilbao, Parque Científico y Tecnológico de Bizkaia, Laida Bidea, Edificio 207-B, 48160 Derio, Bizkaia, Spain}
\affiliation[2]{European Spallation Source, Box 176, SE-221 00 Lund, Sweden}
\affiliation[3]{Center of Excellence for Advanced Materials and Sensing Devices, Ruđer Bo\v{s}kovi\'c Institute, 10000 Zagreb, Croatia}
\affiliation[4]{IPHC, Universit\'{e} de Strasbourg, CNRS/IN2P3, Strasbourg, France}
\affiliation[5]{Department of Physics, School of Engineering Sciences, KTH Royal Institute of Technology,
Roslagstullsbacken 21, 106 91 Stockholm, Sweden}
\affiliation[6]{The Oskar Klein Centre, AlbaNova University Center, Roslagstullsbacken 21, 106 91 Stockholm, Sweden}
\affiliation[7]{Sofia University St. Kliment Ohridski, Faculty of Physics, 1164 Sofia, Bulgaria}
\affiliation[8]{University of Milano-Bicocca and INFN Sez. di Milano-Bicocca, 20126 Milano, Italy}
\affiliation[9]{Department of Physics, Lund University, P.O Box 118, 221 00 Lund, Sweden}
\affiliation[10]{Department of Physics and Astronomy, FREIA Division, Uppsala University, P.O. Box 516, 751 20 Uppsala, Sweden}
\affiliation[11]{Faculty of Engineering, Lund University, P.O Box 118, 221 00 Lund, Sweden}
\affiliation[12]{AGH University of Krakow, al. A. Mickiewicza 30, 30-059 Krakow, Poland}
\affiliation[13]{INFN Sez. di Padova, Padova, Italy}

\affiliation[14]{CERN, 1211 Geneva 23, Switzerland}
\affiliation[15]{Institute of Nuclear and Particle Physics, NCSR Demokritos, Neapoleos 27, 15341 Agia Paraskevi, Greece}
\affiliation[16]{Cockroft Institute (A36), Liverpool University, Warrington WA4 4AD, UK}
\affiliation[17]{Department of Physics, Aristotle University of Thessaloniki, Thessaloniki, Greece}
\affiliation[18]{Center for Interdisciplinary Research and Innovation (CIRI-AUTH), Thessaloniki, Greece}
\affiliation[19]{Institute for Advanced Research, Nagoya University, Nagoya 464–8601, Japan}
\affiliation[20]{Dipartimento di Matematica e Fisica, Universit\'a di Roma Tre, Via della Vasca Navale 84, 00146 Rome, Italy}
\affiliation[21]{University of Cukurova, Faculty of Science and Letters, Department of Physics, 01330 Adana, Turkey}
\affiliation[22]{Department of Physics and Astronomy, Ghent University, Proeftuinstraat 86, B-9000 Ghent, Belgium}
\affiliation[23]{Theory Division, Saha Institute of Nuclear Physics, 1/AF, Bidhannagar, Kolkata 700064, India}
\affiliation[24]{Homi  Bhabha  National  Institute,  Anushakti  Nagar,  Mumbai  400094,  India}
\affiliation[25]{Institute for Experimental Physics, Hamburg University, 22761 Hamburg, Germany}
\affiliation[26]{Physics Laboratory, School of Science and Technology, Hellenic Open University, 26335, Patras, Greece}
\affiliation[27]{Department of Physics and Astronomy "G. Galilei", University of Padova and INFN Sezione di Padova, Italy}
\affiliation[28]{Department of Physics, Ramakrishna Mission Vivekananda Educational and Research Institute, Belur Math, Howrah, West Bengal 711202, India}
\affiliation[29]{Department of Civil, Environmental and Natural Resources Engineering, Luleå University of Technology, SE-971 87 Lulea, Sweden}
\affiliation[30]{Institute of High Energy Physics (IHEP) Dongguan Campus, Chinese Academy of Sciences (CAS), Guangdong 523803, China}

\emailAdd{mghosh@irb.hr}
\emailAdd{alessio.giarnetti@uniroma3.it}
\emailAdd{aman.gupta@saha.ac.in}
\emailAdd{davide.meloni@uniroma3.it}

%% file: main.bbl
\providecommand{\href}[2]{#2}\begingroup\raggedright\begin{thebibliography}{10}

\bibitem{Super-Kamiokande:1998kpq}
{\bf Super-Kamiokande} Collaboration, Y.~Fukuda et~al., {\it {Evidence for oscillation of atmospheric neutrinos}},  {\em Phys. Rev. Lett.} {\bf 81} (1998) 1562--1567, [\href{http://arxiv.org/abs/hep-ex/9807003}{{\tt hep-ex/9807003}}].

\bibitem{SNO:2002tuh}
{\bf SNO} Collaboration, Q.~R. Ahmad et~al., {\it {Direct evidence for neutrino flavor transformation from neutral current interactions in the Sudbury Neutrino Observatory}},  {\em Phys. Rev. Lett.} {\bf 89} (2002) 011301, [\href{http://arxiv.org/abs/nucl-ex/0204008}{{\tt nucl-ex/0204008}}].

\bibitem{Kajita:2016cak}
T.~Kajita, {\it {Nobel Lecture: Discovery of atmospheric neutrino oscillations}},  {\em Rev. Mod. Phys.} {\bf 88} (2016) 030501.

\bibitem{McDonald:2016ixn}
A.~B. McDonald, {\it {Nobel Lecture: The Sudbury Neutrino Observatory: Observation of flavor change for solar neutrinos}},  {\em Rev. Mod. Phys.} {\bf 88} (2016) 030502.

\bibitem{Wolfenstein:1977ue}
L.~Wolfenstein, {\it {Neutrino Oscillations in Matter}},  {\em Phys. Rev. D} {\bf 17} (1978) 2369--2374.

\bibitem{Kopp:2007ne}
J.~Kopp, M.~Lindner, T.~Ota, and J.~Sato, {\it {Non-standard neutrino interactions in reactor and superbeam experiments}},  {\em Phys. Rev. D} {\bf 77} (2008) 013007, [\href{http://arxiv.org/abs/0708.0152}{{\tt arXiv:0708.0152}}].

\bibitem{Du:2020dwr}
Y.~Du, H.-L. Li, J.~Tang, S.~Vihonen, and J.-H. Yu, {\it {Non-standard interactions in SMEFT confronted with terrestrial neutrino experiments}},  {\em JHEP} {\bf 03} (2021) 019, [\href{http://arxiv.org/abs/2011.14292}{{\tt arXiv:2011.14292}}].

\bibitem{Gupta:2023wct}
A.~Gupta, D.~Majumdar, and S.~Prakash, {\it {Neutrino oscillation measurements with KamLAND and JUNO in the presence of scalar NSI}},  \href{http://arxiv.org/abs/2306.07343}{{\tt arXiv:2306.07343}}.

\bibitem{ESSnuSB:2023lbg}
{\bf ESSnuSB} Collaboration, J.~Aguilar et~al., {\it {Study of nonstandard interactions mediated by a scalar field at the ESSnuSB experiment}},  {\em Phys. Rev. D} {\bf 109} (2024), no.~11 115010, [\href{http://arxiv.org/abs/2310.10749}{{\tt arXiv:2310.10749}}].

\bibitem{Denton:2024upc}
P.~B. Denton, A.~Giarnetti, and D.~Meloni, {\it {Solar neutrinos and the strongest oscillation constraints on scalar NSI}},  {\em JHEP} {\bf 01} (2025) 097, [\href{http://arxiv.org/abs/2409.15411}{{\tt arXiv:2409.15411}}].

\bibitem{Ohlsson:2012kf}
T.~Ohlsson, {\it {Status of non-standard neutrino interactions}},  {\em Rept. Prog. Phys.} {\bf 76} (2013) 044201, [\href{http://arxiv.org/abs/1209.2710}{{\tt arXiv:1209.2710}}].

\bibitem{Miranda:2015dra}
O.~G. Miranda and H.~Nunokawa, {\it {Non standard neutrino interactions: current status and future prospects}},  {\em New J. Phys.} {\bf 17} (2015), no.~9 095002, [\href{http://arxiv.org/abs/1505.06254}{{\tt arXiv:1505.06254}}].

\bibitem{Farzan:2017xzy}
Y.~Farzan and M.~Tortola, {\it {Neutrino oscillations and Non-Standard Interactions}},  {\em Front. in Phys.} {\bf 6} (2018) 10, [\href{http://arxiv.org/abs/1710.09360}{{\tt arXiv:1710.09360}}].

\bibitem{10.21468/SciPostPhysProc.2.001}
P.~S. Bhupal~Dev et~al., {\it {Neutrino Non-Standard Interactions: A Status Report}},  {\em SciPost Phys. Proc.} {\bf 2} (2019) 001, [\href{http://arxiv.org/abs/1907.00991}{{\tt arXiv:1907.00991}}].

\bibitem{Huitu:2016bmb}
K.~Huitu, T.~J. K\"arkk\"ainen, J.~Maalampi, and S.~Vihonen, {\it {Constraining the nonstandard interaction parameters in long baseline neutrino experiments}},  {\em Phys. Rev. D} {\bf 93} (2016), no.~5 053016, [\href{http://arxiv.org/abs/1601.07730}{{\tt arXiv:1601.07730}}].

\bibitem{Chaves:2021kxe}
M.~E. Chaves, P.~C. de~Holanda, and O.~L.~G. Peres, {\it {Testing non-standard neutrino interactions in (anti)-electron neutrino disappearance experiments}},  {\em JHEP} {\bf 03} (2023) 180, [\href{http://arxiv.org/abs/2106.15725}{{\tt arXiv:2106.15725}}].

\bibitem{He:1990pn}
X.~G. He, G.~C. Joshi, H.~Lew, and R.~R. Volkas, {\it {NEW Z-prime PHENOMENOLOGY}},  {\em Phys. Rev. D} {\bf 43} (1991) 22--24.

\bibitem{Foot:1990mn}
R.~Foot, {\it {New Physics From Electric Charge Quantization?}},  {\em Mod. Phys. Lett. A} {\bf 6} (1991) 527--530.

\bibitem{He:1991qd}
X.-G. He, G.~C. Joshi, H.~Lew, and R.~R. Volkas, {\it {Simplest Z-prime model}},  {\em Phys. Rev. D} {\bf 44} (1991) 2118--2132.

\bibitem{Foot:1994vd}
R.~Foot, X.~G. He, H.~Lew, and R.~R. Volkas, {\it {Model for a light Z-prime boson}},  {\em Phys. Rev. D} {\bf 50} (1994) 4571--4580, [\href{http://arxiv.org/abs/hep-ph/9401250}{{\tt hep-ph/9401250}}].

\bibitem{Dolgov:1999gk}
A.~D. Dolgov, {\it {Long range forces in the universe}},  {\em Phys. Rept.} {\bf 320} (1999) 1--15.

\bibitem{Smirnov:2019cae}
A.~Y. Smirnov and X.-J. Xu, {\it {Wolfenstein potentials for neutrinos induced by ultra-light mediators}},  {\em JHEP} {\bf 12} (2019) 046, [\href{http://arxiv.org/abs/1909.07505}{{\tt arXiv:1909.07505}}].

\bibitem{Pontecorvo:1967fh}
B.~Pontecorvo, {\it {Neutrino Experiments and the Problem of Conservation of Leptonic Charge}},  {\em Zh. Eksp. Teor. Fiz.} {\bf 53} (1967) 1717--1725.

\bibitem{Gribov:1968kq}
V.~N. Gribov and B.~Pontecorvo, {\it {Neutrino astronomy and lepton charge}},  {\em Phys. Lett. B} {\bf 28} (1969) 493.

\bibitem{Cirigliano:2005ck}
V.~Cirigliano, B.~Grinstein, G.~Isidori, and M.~B. Wise, {\it {Minimal flavor violation in the lepton sector}},  {\em Nucl. Phys. B} {\bf 728} (2005) 121--134, [\href{http://arxiv.org/abs/hep-ph/0507001}{{\tt hep-ph/0507001}}].

\bibitem{Altarelli:2010gt}
G.~Altarelli and F.~Feruglio, {\it {Discrete Flavor Symmetries and Models of Neutrino Mixing}},  {\em Rev. Mod. Phys.} {\bf 82} (2010) 2701--2729, [\href{http://arxiv.org/abs/1002.0211}{{\tt arXiv:1002.0211}}].

\bibitem{Asai:2017ryy}
K.~Asai, K.~Hamaguchi, and N.~Nagata, {\it {Predictions for the neutrino parameters in the minimal gauged U(1)$_{L_\mu-L_\tau}$ model}},  {\em Eur. Phys. J. C} {\bf 77} (2017) 763, [\href{http://arxiv.org/abs/1705.00419}{{\tt arXiv:1705.00419}}].

\bibitem{Asai:2018ocx}
K.~Asai, K.~Hamaguchi, N.~Nagata, S.-Y. Tseng, and K.~Tsumura, {\it {Minimal Gauged U(1)$_{L_\alpha - L_\beta}$ Models Driven into a Corner}},  {\em Phys. Rev. D} {\bf 99} (2019) 055029, [\href{http://arxiv.org/abs/1811.07571}{{\tt arXiv:1811.07571}}].

\bibitem{Lou:2024fvw}
Y.~Lou and T.~Nomura, {\it {Neutrino observables in gauged $U(1)_{L_\alpha-L_\beta}$ models with two Higgs doublet and one singlet scalars}},  \href{http://arxiv.org/abs/2406.01030}{{\tt arXiv:2406.01030}}.

\bibitem{Grifols:2003gy}
J.~A. Grifols and E.~Masso, {\it {Neutrino oscillations in the sun probe long range leptonic forces}},  {\em Phys. Lett. B} {\bf 579} (2004) 123--126, [\href{http://arxiv.org/abs/hep-ph/0311141}{{\tt hep-ph/0311141}}].

\bibitem{Bandyopadhyay:2006uh}
A.~Bandyopadhyay, A.~Dighe, and A.~S. Joshipura, {\it {Constraints on flavor-dependent long range forces from solar neutrinos and KamLAND}},  {\em Phys. Rev. D} {\bf 75} (2007) 093005, [\href{http://arxiv.org/abs/hep-ph/0610263}{{\tt hep-ph/0610263}}].

\bibitem{Gonzalez-Garcia:2006vic}
M.~C. Gonzalez-Garcia, P.~C. de~Holanda, E.~Masso, and R.~Zukanovich~Funchal, {\it {Probing long-range leptonic forces with solar and reactor neutrinos}},  {\em JCAP} {\bf 01} (2007) 005, [\href{http://arxiv.org/abs/hep-ph/0609094}{{\tt hep-ph/0609094}}].

\bibitem{Joshipura:2003jh}
A.~S. Joshipura and S.~Mohanty, {\it {Constraints on flavor dependent long range forces from atmospheric neutrino observations at super-Kamiokande}},  {\em Phys. Lett. B} {\bf 584} (2004) 103--108, [\href{http://arxiv.org/abs/hep-ph/0310210}{{\tt hep-ph/0310210}}].

\bibitem{Bustamante:2018mzu}
M.~Bustamante and S.~K. Agarwalla, {\it {Universe's Worth of Electrons to Probe Long-Range Interactions of High-Energy Astrophysical Neutrinos}},  {\em Phys. Rev. Lett.} {\bf 122} (2019) 061103, [\href{http://arxiv.org/abs/1808.02042}{{\tt arXiv:1808.02042}}].

\bibitem{Agarwalla:2023sng}
S.~K. Agarwalla, M.~Bustamante, S.~Das, and A.~Narang, {\it {Present and future constraints on flavor-dependent long-range interactions of high-energy astrophysical neutrinos}},  {\em JHEP} {\bf 08} (2023) 113, [\href{http://arxiv.org/abs/2305.03675}{{\tt arXiv:2305.03675}}].

\bibitem{Coloma:2020gfv}
P.~Coloma, M.~C. Gonzalez-Garcia, and M.~Maltoni, {\it {Neutrino oscillation constraints on U(1)' models: from non-standard interactions to long-range forces}},  {\em JHEP} {\bf 01} (2021) 114, [\href{http://arxiv.org/abs/2009.14220}{{\tt arXiv:2009.14220}}]. [Erratum: \textit{JHEP} \textbf{11} (2022) 115].

\bibitem{Chatterjee:2015gta}
S.~S. Chatterjee, A.~Dasgupta, and S.~K. Agarwalla, {\it {Exploring Flavor-Dependent Long-Range Forces in Long-Baseline Neutrino Oscillation Experiments}},  {\em JHEP} {\bf 12} (2015) 167, [\href{http://arxiv.org/abs/1509.03517}{{\tt arXiv:1509.03517}}].

\bibitem{Khatun:2018lzs}
A.~Khatun, T.~Thakore, and S.~Kumar~Agarwalla, {\it {Can INO be Sensitive to Flavor-Dependent Long-Range Forces?}},  {\em JHEP} {\bf 04} (2018) 023, [\href{http://arxiv.org/abs/1801.00949}{{\tt arXiv:1801.00949}}].

\bibitem{Singh:2023nek}
M.~Singh, M.~Bustamante, and S.~K. Agarwalla, {\it {Flavor-dependent long-range neutrino interactions in DUNE \& T2HK: alone they constrain, together they discover}},  {\em JHEP} {\bf 08} (2023) 101, [\href{http://arxiv.org/abs/2305.05184}{{\tt arXiv:2305.05184}}].

\bibitem{Mishra:2024riq}
P.~Mishra, R.~Majhi, S.~K. Pusty, M.~Ghosh, and R.~Mohanta, {\it {Study of long range force in P2SO and T2HKK}},  {\em JHEP} {\bf 09} (2024) 100, [\href{http://arxiv.org/abs/2402.19178}{{\tt arXiv:2402.19178}}].

\bibitem{T2K:2017hed}
{\bf T2K} Collaboration, K.~Abe et~al., {\it {Combined Analysis of Neutrino and Antineutrino Oscillations at T2K}},  {\em Phys. Rev. Lett.} {\bf 118} (2017) 151801, [\href{http://arxiv.org/abs/1701.00432}{{\tt arXiv:1701.00432}}].

\bibitem{T2K:2019bcf}
{\bf T2K} Collaboration, K.~Abe et~al., {\it {Constraint on the matter\textendash{}antimatter symmetry-violating phase in neutrino oscillations}},  {\em Nature} {\bf 580} (2020) 339--344, [\href{http://arxiv.org/abs/1910.03887}{{\tt arXiv:1910.03887}}]. [Erratum: \textit{Nature} \textbf{583} (2020) E16].

\bibitem{NOvA:2019cyt}
{\bf NOvA} Collaboration, M.~A. Acero et~al., {\it {First Measurement of Neutrino Oscillation Parameters using Neutrinos and Antineutrinos by NOvA}},  {\em Phys. Rev. Lett.} {\bf 123} (2019), no.~15 151803, [\href{http://arxiv.org/abs/1906.04907}{{\tt arXiv:1906.04907}}].

\bibitem{DUNE:2020jqi}
{\bf DUNE} Collaboration, B.~Abi et~al., {\it {Long-baseline neutrino oscillation physics potential of the DUNE experiment}},  {\em Eur. Phys. J. C} {\bf 80} (2020) 978, [\href{http://arxiv.org/abs/2006.16043}{{\tt arXiv:2006.16043}}].

\bibitem{ESSnuSB:2021azq}
{\bf ESSnuSB} Collaboration, A.~Alekou et~al., {\it {Updated physics performance of the ESSnuSB experiment: ESSnuSB collaboration}},  {\em Eur. Phys. J. C} {\bf 81} (2021) 1130, [\href{http://arxiv.org/abs/2107.07585}{{\tt arXiv:2107.07585}}].

\bibitem{Abele:2022iml}
H.~Abele et~al., {\it {Particle Physics at the European Spallation Source}},  {\em Phys. Rept.} {\bf 1023} (2023) 1--84, [\href{http://arxiv.org/abs/2211.10396}{{\tt arXiv:2211.10396}}].

\bibitem{Alekou:2022emd}
A.~Alekou et~al., {\it {The European Spallation Source neutrino super-beam conceptual design report}},  {\em Eur. Phys. J. ST} {\bf 231} (2022) 3779--3955, [\href{http://arxiv.org/abs/2206.01208}{{\tt arXiv:2206.01208}}]. [Erratum: \textit{Eur. Phys. J. ST} \textbf{232} (2023) 15--16].

\bibitem{ESSnuSB:2023ogw}
{\bf ESSnuSB} Collaboration, A.~Alekou et~al., {\it {The ESSnuSB Design Study: Overview and Future Prospects}},  {\em Universe} {\bf 9} (2023), no.~8 347, [\href{http://arxiv.org/abs/2303.17356}{{\tt arXiv:2303.17356}}].

\bibitem{ESSnuSB:2024tmn}
{\bf ESSnuSB} Collaboration, J.~Aguilar et~al., {\it {Search for Leptonic CP Violation with the ESSnuSBplus Project}},  {\em LHEP} {\bf 2024} (2024) 517.

\bibitem{Agarwalla:2024ylc}
S.~K. Agarwalla, M.~Bustamante, M.~Singh, and P.~Swain, {\it {A plethora of long-range neutrino interactions probed by DUNE and T2HK}},  {\em JHEP} {\bf 09} (2024) 055, [\href{http://arxiv.org/abs/2404.02775}{{\tt arXiv:2404.02775}}].

\bibitem{Choubey:2024krp}
S.~Choubey, S.~Khan, M.~Merchand, and S.~Vihonen, {\it {Constraining dark matter from strong phase transitions in a $ \textrm{U}{(1)}_{L_{\mu }-{L}_{\tau }} $ model: implications for neutrino masses and muon g \ensuremath{-} 2}},  {\em JHEP} {\bf 10} (2024) 186, [\href{http://arxiv.org/abs/2406.16460}{{\tt arXiv:2406.16460}}].

\bibitem{Ibe:2025rwk}
M.~Ibe, S.~Shirai, and K.~Watanabe, {\it {Global Neutrino Constraints on the Minimal U(1)$_{L_\mu-L_\tau}$ Model}},  \href{http://arxiv.org/abs/2503.01399}{{\tt arXiv:2503.01399}}.

\bibitem{Babu:1997st}
K.~S. Babu, C.~F. Kolda, and J.~March-Russell, {\it {Implications of generalized Z - Z-prime mixing}},  {\em Phys. Rev. D} {\bf 57} (1998) 6788--6792, [\href{http://arxiv.org/abs/hep-ph/9710441}{{\tt hep-ph/9710441}}].

\bibitem{Joshipura:2019qxz}
A.~S. Joshipura, N.~Mahajan, and K.~M. Patel, {\it {Generalised $\mu$-$\tau$ symmetries and calculable gauge kinetic and mass mixing in $\mathrm{U}{(1)}_{L_{\mu }-{L}_{\tau }}$ models}},  {\em JHEP} {\bf 03} (2020) 001, [\href{http://arxiv.org/abs/1909.02331}{{\tt arXiv:1909.02331}}].

\bibitem{Holdom:1985ag}
B.~Holdom, {\it {Two U(1)'s and Epsilon Charge Shifts}},  {\em Phys. Lett. B} {\bf 166} (1986) 196--198.

\bibitem{Ghoshal:2020hyo}
A.~Ghoshal, A.~Giarnetti, and D.~Meloni, {\it {Neutrino Invisible Decay at DUNE: a multi-channel analysis}},  {\em J. Phys. G} {\bf 48} (2021), no.~5 055004, [\href{http://arxiv.org/abs/2003.09012}{{\tt arXiv:2003.09012}}].

\bibitem{Gehrlein:2024vwz}
J.~Gehrlein, P.~A.~N. Machado, and J.~a.~P. Pinheiro, {\it {Constraining non-standard neutrino interactions with neutral current events at long-baseline oscillation experiments}},  {\em JHEP} {\bf 05} (2025) 065, [\href{http://arxiv.org/abs/2412.08712}{{\tt arXiv:2412.08712}}].

\bibitem{Denton:2022pxt}
P.~B. Denton, A.~Giarnetti, and D.~Meloni, {\it {How to identify different new neutrino oscillation physics scenarios at DUNE}},  {\em JHEP} {\bf 02} (2023) 210, [\href{http://arxiv.org/abs/2210.00109}{{\tt arXiv:2210.00109}}].

\bibitem{Agarwalla:2021owd}
S.~K. Agarwalla, S.~Das, A.~Giarnetti, and D.~Meloni, {\it {Model-independent constraints on non-unitary neutrino mixing from high-precision long-baseline experiments}},  {\em JHEP} {\bf 07} (2022) 121, [\href{http://arxiv.org/abs/2111.00329}{{\tt arXiv:2111.00329}}].

\bibitem{Coloma:2017ptb}
P.~Coloma, D.~V. Forero, and S.~J. Parke, {\it {DUNE Sensitivities to the Mixing between Sterile and Tau Neutrinos}},  {\em JHEP} {\bf 07} (2018) 079, [\href{http://arxiv.org/abs/1707.05348}{{\tt arXiv:1707.05348}}].

\bibitem{Giarnetti:2021wur}
A.~Giarnetti and D.~Meloni, {\it {New Sources of Leptonic CP Violation at the DUNE Neutrino Experiment}},  {\em Universe} {\bf 7} (2021), no.~7 240, [\href{http://arxiv.org/abs/2106.00030}{{\tt arXiv:2106.00030}}].

\bibitem{Giarnetti:2024mdt}
A.~Giarnetti, S.~Marciano, and D.~Meloni, {\it {Exploring New Physics with Deep Underground Neutrino Experiment High-Energy Flux: The Case of Lorentz Invariance Violation, Large Extra Dimensions and Long-Range Forces}},  {\em Universe} {\bf 10} (2024) 357, [\href{http://arxiv.org/abs/2407.17247}{{\tt arXiv:2407.17247}}].

\bibitem{Berryman:2016szd}
J.~M. Berryman, A.~de~Gouv\^ea, K.~J. Kelly, O.~L.~G. Peres, and Z.~Tabrizi, {\it {Large, Extra Dimensions at the Deep Underground Neutrino Experiment}},  {\em Phys. Rev. D} {\bf 94} (2016), no.~3 033006, [\href{http://arxiv.org/abs/1603.00018}{{\tt arXiv:1603.00018}}].

\bibitem{Wise:2018rnb}
M.~B. Wise and Y.~Zhang, {\it {Lepton Flavorful Fifth Force and Depth-dependent Neutrino Matter Interactions}},  {\em JHEP} {\bf 06} (2018) 053, [\href{http://arxiv.org/abs/1803.00591}{{\tt arXiv:1803.00591}}].

\bibitem{Heeck:2010pg}
J.~Heeck and W.~Rodejohann, {\it {Gauged $L_\mu - L_\tau$ and different Muon Neutrino and Anti-Neutrino Oscillations: MINOS and beyond}},  {\em J. Phys. G} {\bf 38} (2011) 085005, [\href{http://arxiv.org/abs/1007.2655}{{\tt arXiv:1007.2655}}].

\bibitem{Agarwalla:2021zfr}
S.~K. Agarwalla, S.~Das, M.~Masud, and P.~Swain, {\it {Evolution of neutrino mass-mixing parameters in matter with non-standard interactions}},  {\em JHEP} {\bf 11} (2021) 094, [\href{http://arxiv.org/abs/2103.13431}{{\tt arXiv:2103.13431}}].

\bibitem{Kikuchi:2008vq}
T.~Kikuchi, H.~Minakata, and S.~Uchinami, {\it {Perturbation Theory of Neutrino Oscillation with Nonstandard Neutrino Interactions}},  {\em JHEP} {\bf 03} (2009) 114, [\href{http://arxiv.org/abs/0809.3312}{{\tt arXiv:0809.3312}}].

\bibitem{Huber:2004ka}
P.~Huber, M.~Lindner, and W.~Winter, {\it {Simulation of long-baseline neutrino oscillation experiments with GLoBES (General Long Baseline Experiment Simulator)}},  {\em Comput. Phys. Commun.} {\bf 167} (2005) 195, [\href{http://arxiv.org/abs/hep-ph/0407333}{{\tt hep-ph/0407333}}].

\bibitem{Huber:2007ji}
P.~Huber, J.~Kopp, M.~Lindner, M.~Rolinec, and W.~Winter, {\it {New features in the simulation of neutrino oscillation experiments with GLoBES 3.0: General Long Baseline Experiment Simulator}},  {\em Comput. Phys. Commun.} {\bf 177} (2007) 432--438, [\href{http://arxiv.org/abs/hep-ph/0701187}{{\tt hep-ph/0701187}}].

\bibitem{ESSnuSB:2024yji}
{\bf ESSnuSB} Collaboration, J.~Aguilar et~al., {\it {Decoherence in neutrino oscillation at the ESSnuSB experiment}},  {\em JHEP} {\bf 08} (2024) 063, [\href{http://arxiv.org/abs/2404.17559}{{\tt arXiv:2404.17559}}].

\bibitem{Esteban:2020cvm}
I.~Esteban, M.~C. Gonzalez-Garcia, M.~Maltoni, T.~Schwetz, and A.~Zhou, {\it {The fate of hints: updated global analysis of three-flavor neutrino oscillations}},  {\em JHEP} {\bf 09} (2020) 178, [\href{http://arxiv.org/abs/2007.14792}{{\tt arXiv:2007.14792}}].

\bibitem{NuFIT5.2}
{NuFIT 5.2, (2022), \url{http://www.nu-fit.org/}}.

\bibitem{Capozzi:2017ipn}
F.~Capozzi, E.~Di~Valentino, E.~Lisi, A.~Marrone, A.~Melchiorri, and A.~Palazzo, {\it {Global constraints on absolute neutrino masses and their ordering}},  {\em Phys. Rev. D} {\bf 95} (2017) 096014, [\href{http://arxiv.org/abs/2003.08511}{{\tt arXiv:2003.08511}}]. [Addendum: \textit{Phys. Rev. D} \textbf{101} (2020) 116013].

\bibitem{Esteban:2018azc}
I.~Esteban, M.~C. Gonzalez-Garcia, A.~Hernandez-Cabezudo, M.~Maltoni, and T.~Schwetz, {\it {Global analysis of three-flavour neutrino oscillations: synergies and tensions in the determination of $\theta_{23}$, $\delta_{CP}$, and the mass ordering}},  {\em JHEP} {\bf 01} (2019) 106, [\href{http://arxiv.org/abs/1811.05487}{{\tt arXiv:1811.05487}}].

\bibitem{deSalas:2020pgw}
P.~F. de~Salas, D.~V. Forero, S.~Gariazzo, P.~Mart\'\i{}nez-Mirav\'e, O.~Mena, C.~A. Ternes, M.~T\'ortola, and J.~W.~F. Valle, {\it {2020 global reassessment of the neutrino oscillation picture}},  {\em JHEP} {\bf 02} (2021) 071, [\href{http://arxiv.org/abs/2006.11237}{{\tt arXiv:2006.11237}}].

\bibitem{Capozzi:2021fjo}
F.~Capozzi, E.~Di~Valentino, E.~Lisi, A.~Marrone, A.~Melchiorri, and A.~Palazzo, {\it {Unfinished fabric of the three neutrino paradigm}},  {\em Phys. Rev. D} {\bf 104} (2021) 083031, [\href{http://arxiv.org/abs/2107.00532}{{\tt arXiv:2107.00532}}].

\bibitem{Gonzalez-Garcia:2021dve}
M.~C. Gonzalez-Garcia, M.~Maltoni, and T.~Schwetz, {\it {NuFIT: Three-Flavour Global Analyses of Neutrino Oscillation Experiments}},  {\em Universe} {\bf 7} (2021) 459, [\href{http://arxiv.org/abs/2111.03086}{{\tt arXiv:2111.03086}}].

\bibitem{Jiang:2024viw}
J.-Q. Jiang, W.~Giar\`e, S.~Gariazzo, M.~G. Dainotti, E.~Di~Valentino, O.~Mena, D.~Pedrotti, S.~S. da~Costa, and S.~Vagnozzi, {\it {Neutrino cosmology after DESI: tightest mass upper limits, preference for the normal ordering, and tension with terrestrial observations}},  \href{http://arxiv.org/abs/2407.18047}{{\tt arXiv:2407.18047}}.

\bibitem{Huber:2002mx}
P.~Huber, M.~Lindner, and W.~Winter, {\it {Superbeams versus neutrino factories}},  {\em Nucl. Phys. B} {\bf 645} (2002) 3--48, [\href{http://arxiv.org/abs/hep-ph/0204352}{{\tt hep-ph/0204352}}].

\bibitem{Fogli:2002pt}
G.~L. Fogli, E.~Lisi, A.~Marrone, D.~Montanino, and A.~Palazzo, {\it {Getting the most from the statistical analysis of solar neutrino oscillations}},  {\em Phys. Rev. D} {\bf 66} (2002) 053010, [\href{http://arxiv.org/abs/hep-ph/0206162}{{\tt hep-ph/0206162}}].

\bibitem{Wilks:1938dza}
S.~S. Wilks, {\it {The Large-Sample Distribution of the Likelihood Ratio for Testing Composite Hypotheses}},  {\em Annals Math. Statist.} {\bf 9} (1938), no.~1 60--62.

\bibitem{Dehnen:1996fa}
W.~Dehnen and J.~Binney, {\it {Mass models of the Milky Way}},  {\em Mon. Not. Roy. Astron. Soc.} {\bf 294} (1998) 429, [\href{http://arxiv.org/abs/astro-ph/9612059}{{\tt astro-ph/9612059}}].

\bibitem{Miller:2013nza}
M.~J. Miller and J.~N. Bregman, {\it {The Structure of the Milky Way's Hot Gas Halo}},  {\em Astrophys. J.} {\bf 770} (2013) 118, [\href{http://arxiv.org/abs/1305.2430}{{\tt arXiv:1305.2430}}].

\bibitem{Baryakhtar:2017ngi}
M.~Baryakhtar, R.~Lasenby, and M.~Teo, {\it {Black Hole Superradiance Signatures of Ultralight Vectors}},  {\em Phys. Rev. D} {\bf 96} (2017) 035019, [\href{http://arxiv.org/abs/1704.05081}{{\tt arXiv:1704.05081}}].

\bibitem{Arkani-Hamed:2006emk}
N.~Arkani-Hamed, L.~Motl, A.~Nicolis, and C.~Vafa, {\it {The String landscape, black holes and gravity as the weakest force}},  {\em JHEP} {\bf 06} (2007) 060, [\href{http://arxiv.org/abs/hep-th/0601001}{{\tt hep-th/0601001}}].

\bibitem{Agarwalla:2022xdo}
S.~K. Agarwalla, S.~Das, A.~Giarnetti, D.~Meloni, and M.~Singh, {\it {Enhancing sensitivity to leptonic CP violation using complementarity among DUNE, T2HK, and T2HKK}},  {\em Eur. Phys. J. C} {\bf 83} (2023) 694, [\href{http://arxiv.org/abs/2211.10620}{{\tt arXiv:2211.10620}}].

\bibitem{Davoudiasl:2011sz}
H.~Davoudiasl, H.-S. Lee, and W.~J. Marciano, {\it {Long-Range Lepton Flavor Interactions and Neutrino Oscillations}},  {\em Phys. Rev. D} {\bf 84} (2011) 013009, [\href{http://arxiv.org/abs/1102.5352}{{\tt arXiv:1102.5352}}].

\bibitem{Araki:2012ip}
T.~Araki, J.~Heeck, and J.~Kubo, {\it {Vanishing Minors in the Neutrino Mass Matrix from Abelian Gauge Symmetries}},  {\em JHEP} {\bf 07} (2012) 083, [\href{http://arxiv.org/abs/1203.4951}{{\tt arXiv:1203.4951}}].

\bibitem{delaVega:2021wpx}
L.~M.~G. de~la Vega, L.~J. Flores, N.~Nath, and E.~Peinado, {\it {Complementarity between dark matter direct searches and CE\ensuremath{\nu}NS experiments in U(1)' models}},  {\em JHEP} {\bf 09} (2021) 146, [\href{http://arxiv.org/abs/2107.04037}{{\tt arXiv:2107.04037}}].

\bibitem{Farzan:2016wym}
Y.~Farzan and J.~Heeck, {\it {Neutrinophilic nonstandard interactions}},  {\em Phys. Rev. D} {\bf 94} (2016) 053010, [\href{http://arxiv.org/abs/1607.07616}{{\tt arXiv:1607.07616}}].

\bibitem{Almumin:2022rml}
Y.~Almumin, M.-C. Chen, M.~Cheng, V.~Knapp-Perez, Y.~Li, A.~Mondol, S.~Ramos-Sanchez, M.~Ratz, and S.~Shukla, {\it {Neutrino Flavor Model Building and the Origins of Flavor and CP Violation}},  {\em Universe} {\bf 9} (2023) 512, [\href{http://arxiv.org/abs/2204.08668}{{\tt arXiv:2204.08668}}].

\end{thebibliography}\endgroup
